\documentclass[10pt,twocolumn]{article}
\usepackage{times}

\usepackage{fancyhdr,graphicx,amsmath,amssymb}
\usepackage{algorithmic}
\usepackage{comment}
\usepackage{subcaption}

\usepackage{hyperref}
\usepackage{cleveref}

\begin{document}

\title{Akita: A CPU scheduler for virtualized Clouds}

\author{Esmail Asyabi$^1$, Azer Bestavros$^1$, Renato Mancuso$^1$, Richard West$^1$, Erfan Sharafzadeh$^2$ \\
$^1$ Boston University  \hspace{10mm}  $^2$Johns Hopkins University \\
$^1$\{easyabi, best, rmancuso, richwest\}@bu.edu  \hspace{10mm}      $^2$erfan@cs.jhu.edu
}
\date{}
\maketitle

\begin{abstract}
Clouds inherit CPU scheduling policies of operating systems. These policies enforce fairness while leveraging  best-effort mechanisms to enhance responsiveness of all schedulable entities, irrespective of their service level objectives (SLOs). This leads to  unpredictable performance that forces cloud providers  to enforce strict reservation and isolation policies to prevent high-criticality services (e.g., Memcached) from being impacted by low-criticality ones (e.g., logging), which results in low utilization. 
    
In this paper, we present Akita, a hypervisor CPU scheduler that delivers predictable performance at high utilization. Akita allows virtual machines (VMs) to be categorized into high- and low-criticality VMs. Akita provides strong guarantees on the ability of cloud providers to meet  SLOs of high-criticality VMs, by temporarily slowing down low-criticality VMs if necessary. Akita, therefore, allows the co-existence of high and low-criticality VMs on the same physical machine, leading to higher utilization. The effectiveness of Akita is demonstrated by a prototype implementation in the Xen hypervisor. We present experimental results that show the many advantages of adopting Akita as the hypervisor CPU scheduler. In particular, we show that high-criticality Memcached VMs are able to deliver predictable performance despite being co-located with low-criticality CPU-bound VMs.
\end{abstract} 

\section{Introduction}
 To be able to harness the economies of scale, data centers tend to consolidate  diverse workloads ranging from critical (e.g., in-memory key-value stores) to non-critical services (e.g.,  logging, batch or video processing) onto as few physical machines (PMs) as possible, to raise utilization, mitigate operational costs, and/or reduce energy consumption \cite{consolidation} \cite{Elasticity} \cite{PerfIso} \cite{Shenango} \cite{Arachne}. In virtualized clouds, however,  many research studies show  that under policies currently adopted for   processor sharing, co-located workloads introduce a level of unpredictability regarding the performance of critical applications \cite{Reconciling} \cite{consolidation} \cite{interference} \cite{isolation} \cite{Jang2015-gf} \cite{terrierTail}. Therefore, to deliver the expected Quality of Service (QoS), cloud providers resort to strict isolation and reservation policies, leading to significant inefficiencies resulting from wasted/unused processor capacities \cite{facbook-util} \cite{utune} \cite{Shenango} \cite{PerfIso}.  

Existing CPU scheduling policies  currently adopted in clouds (e.g., Xen's Credit schedulers and Linux's CFS scheduler) are inherited from operating systems.  The main goal of these policies is to  enforce fair allocation of CPU shares which are determined by virtual CPU (vCPU) budgets \cite{Zhuravlev2012-vj}. For I/O, however, they rely on best-effort approaches to raise responsiveness of all IO-bound vCPUs by reducing their CPU Access Latencies (CALs) as much as possible. For example, Xen's credit scheduler employs a load balancer to exploit all existing cores in order to reduce CALs of IO-bound vCPUs  \cite{cts} \cite{terrierTail}. 

Although best-effort  policies improve  responsiveness, we argue that they impose several challenges to virtualized clouds: Best-effort mechanisms try to raise responsiveness of all vCPUs/VMs irrespective of their Service Level Objectives (SLOs). This  contradicts  the cloud \textit{pay-as-you-go}  business model since these policies might utilize all processor resources (through load-balancing) to deliver the IO performance that is neither expected nor  paid for. More importantly,  these policies lead to unpredictability, where  delivered IO performance of a virtual machine (VM) depends on the behavior/existence of other VMs, making it very difficult to predict VMs' delivered QoS in advance. This forces cloud providers to resort to costly middleware in order to monitor QoS of VMs to prevent  SLO violations (e.g., by migrating VMs \cite{consolidation}), or  to seek predictability through over-provisioning (e.g., by pining  each  vCPU to a distinct CPU core),  resulting in lower utilization,  more  power consumption and higher operational costs  \cite{facbook-util} \cite{Reconciling} \cite{Bobtail}. 

 We present Akita, a CPU scheduler for  virtualized  clouds. Akita offers several advantages to  cloud providers. First, it offers a predictable IO performance even at high utilization of  processor resources. Second, it  characterises VMs/vCPUs with not only a CPU budget but also an IO Quality (IOQ). Adopting this information, Akita scheduler offers differentiated service qualities regarding both execution time and IO performance of VMs determined by VMs' budgets and IOQs, respectively. Third, unlike existing schedulers, Akita operates  the lowest possible number of CPU cores while delivering the expected QoS. Finally,  Akita offers a schedulability test that lets cloud providers know if a PM can accommodate a VM's IO and CPU requirements  in advance, mitigating the need for costly    monitoring services in cloud data centers.  
 
To achieve these goals, in Akita, we characterize VMs/vCPUs with four attributes, a pessimistic budget, an optimistic budget, an IO quality metric and a  critically level.  A VM's optimistic budget  indicates its  average required CPU time, the pessimistic budget shows its worst case required CPU time. IO quality indicates the average response time tolerated by the VM, and the critically level which is either high or low indicates if the VM is running a critical (e.g., a user facing) application or a non-critical (e.g., background or free service) application.   
 
 In each CPU core, Akita scheduler alternates between two modes: normal and critical modes. When mode is normal, Akita allocates all vCPUs the CPU shares determined by their optimistic  budgets with  CPU access latencies that are not higher than their IO qualities. Meanwhile, Akita monitors  high-criticality vCPUs. If a high-criticality vCPU requires its pessimistic CPU budget, Akita changes its mode to the high critical mode. In the the high critical mode,  Akita ignores low-criticality vCPUs to accommodate   required CPU shares of high-criticality vCPUs.  When high-criticality vCPUs no longer need their pessimistic budgets, Akita resets its mode to normal mode again. Therefore,  (1) Akita increases  utilization of processor resources by consolidating high and low-criticality vCPUs on the same pCPU (2) enforces fairness by allocating VMs CPU shares corresponding to VMs' budgets (3) offers different  service qualities for IO, by keeping the CALs of vCPUs less than their defined IOQs, and (4) finally offers a  predictable performance for high-criticality VMs by temporarily slowing down/ignore low criticality  ones if needed. 
 
Further, Akita distributes vCPUs based on a first-fit approach. When a new vCPU is created, Akita finds the first pCPU that can accommodate the vCPUs requirements in terms of both CPU share and IO quality. Therefore, Akita operates as few CPU cores as possible while delivering the expected performance as apposed to existing schedulers that utilize all CPU cores. In summary, our contributions are as follows:
\begin{itemize}
\item  We show that existing CPU schedulers adopted in virtualized clouds lead to unpredictability and consequently low utilization. 
\item  We present  Akita, a hypervisor CPU scheduler that raises utilization while offering a  predictable  IO performance.

\item  We have built a prototype  of Akita in the Xen hypervisor. We present experimental results that show the many advantages of Akita as the hypervisor CPU scheduler. In particular, we show that HI-crit Memcached/RPC VMs are able to deliver predictable performance despite being co-located with LO-crit CPU-bound VMs.
\end{itemize}

\section{Motivation and Background}  \label{ssec:motiv}
In this section, we study the essential properties of a hypervisor CPU scheduler that improves  performance predictability and utilization of  virtualized  clouds. We then discuss why existing CPU scheduling policies are not able to address the hard problem of predictable IO at high utilization. 

\subsection{ The key properties}
\textbf{ Fairness.}  The scheduler must enforce fairness where CPU shares are allocated corresponding to VMs budgets. These budgets  are determined by the cloud provider based on SLOs and VM prices, offering different execution time to CPU bound workloads.

\textbf{Differentiated IO qualities.} Similar to CPU shares, we argue that the scheduler must also offer different   IO qualities, ranging from best effort to guaranteed and predictable IO.

\textbf{High utilization.} Predictability is easily achievable through over-provisioning. Therefore, we argue that  predictable IO  should be offered at high utilization of processor resources, which in turn cuts  operational  and power costs.

\textbf{Schedulability test.}    The scheduler must let the cloud providers know if a PM can accommodate a VM's requirements (CPU share and IO quality) in advance. This mitigates the need  for costly monitoring services. 
\subsection{Existing schedulers}

\begin{figure}
\begin{subfigure}{0.44\linewidth}

  \includegraphics[width= 4cm,height=3cm, keepaspectratio]{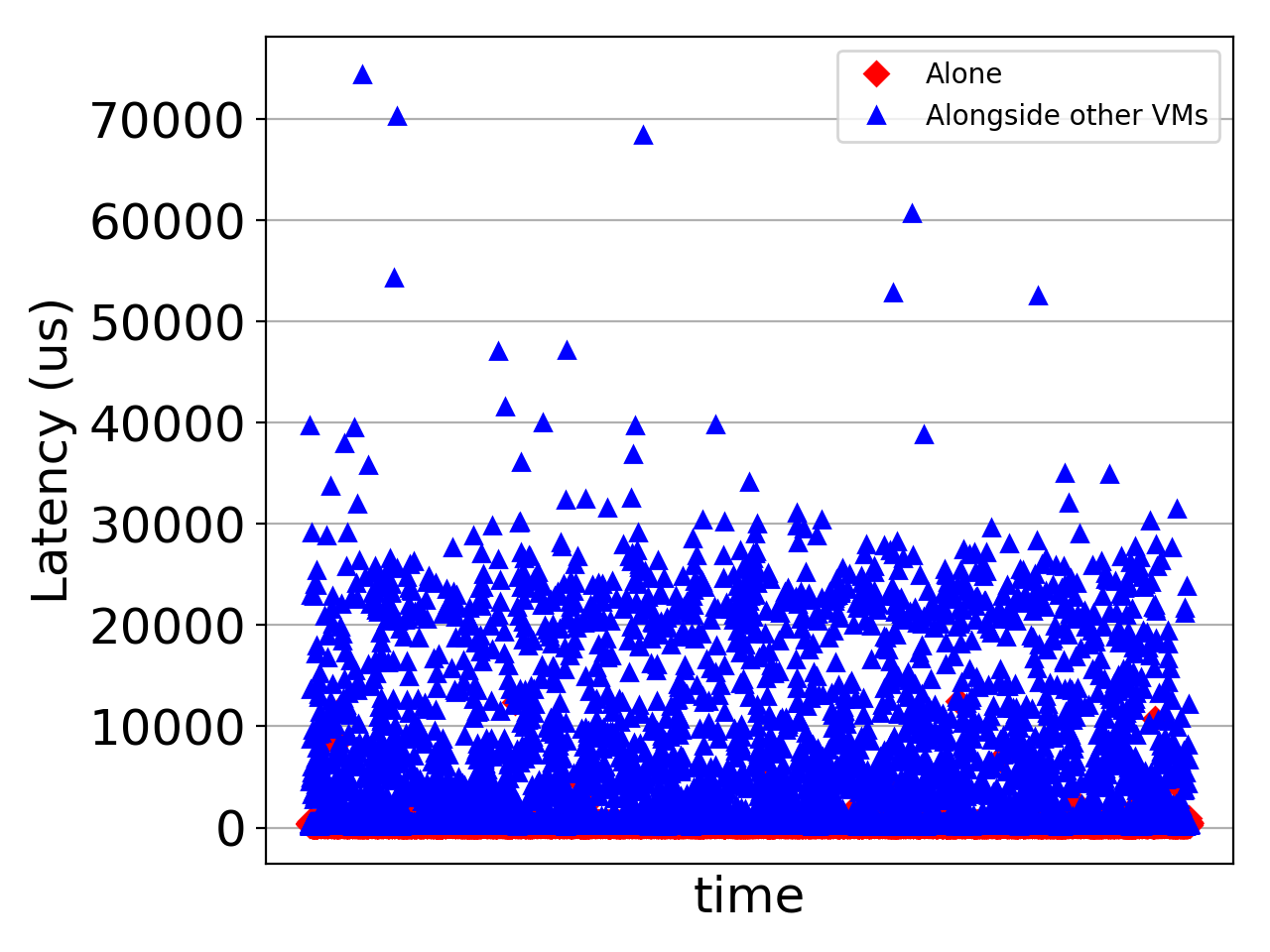}
  \caption{ Xen}
  \label{motiv_xen}
 \end{subfigure}
 ~
\begin{subfigure}{0.42\linewidth }
 
  \includegraphics[width=4cm,height=3cm, keepaspectratio]{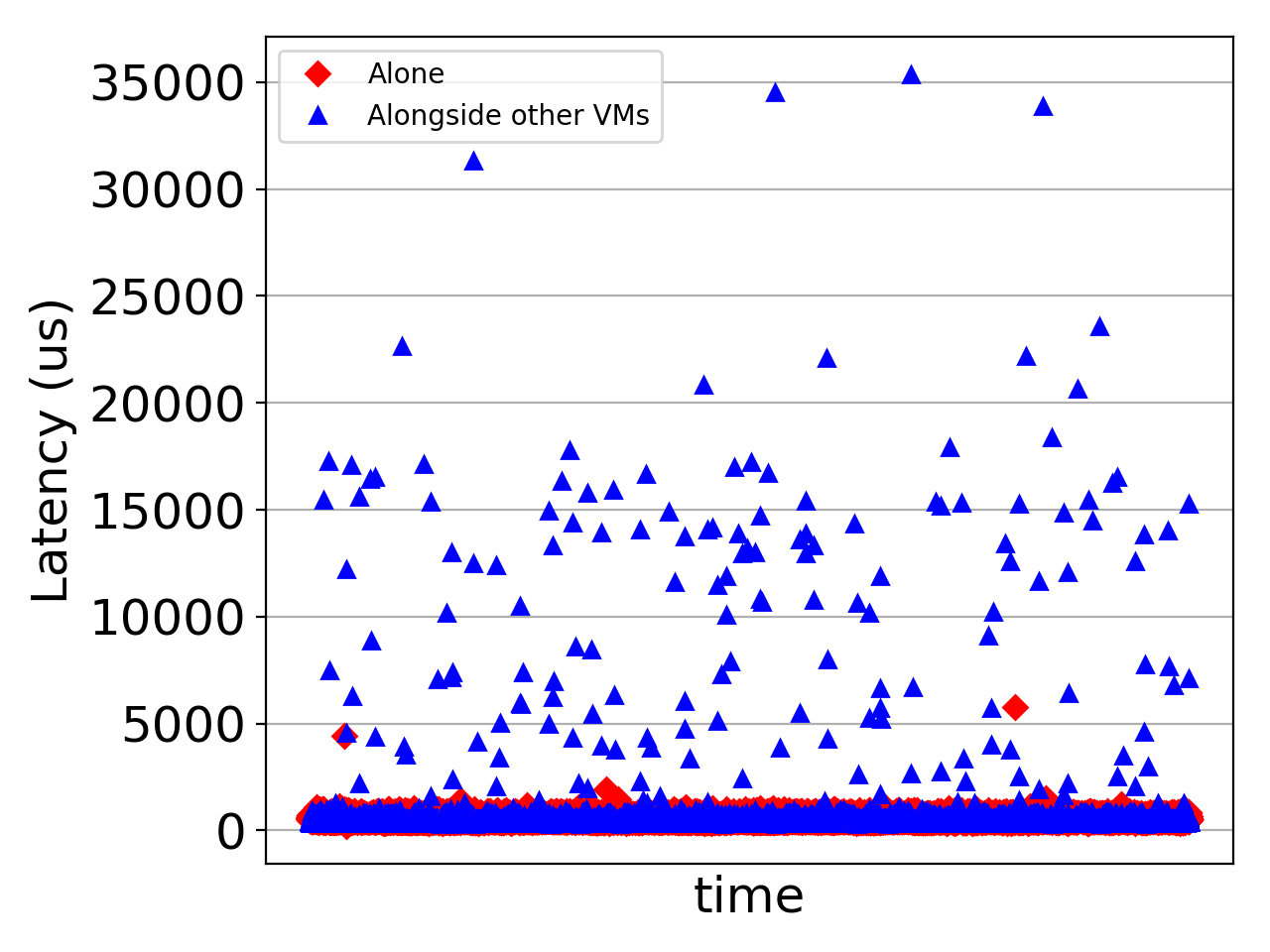}
  \caption{KVM}
  \label{motiv_kvm}
\end{subfigure}

\caption{ RTTs of Memcached requests }
\label{motiv}
\end{figure}

\textbf{The Credit scheduler of Xen}  uses a  proportional weighting policy to share processor resources among executable entities (vCPUs). It assigns each vCPU a relative weight that determines the amount of CPU time that can be allocated to a vCPU relative to other vCPUs.  The scheduler serves vCPUs in round-robin fashion with the time slice of 30ms. Its main goal is to share processor resources fairly augmented by a best-effort approach (boosting) to increase the performance of IO-bound vCPUs by giving the highest priority to the newly awakened vCPUs which are very likely to be IO-bound vCPUs. The IO performance is further improved by a load balancer that reduces the waiting time of vCPUs by migrating vCPUs from the queues of busier CPUs to idle CPUs.  Therefore, by adjusting the weight, different CPU shares can be allocated to VMs, enforcing fairness. When it comes to IO performance,  although  Xen's best-effort mechanisms (boosting and load balancing)  improve the performance of  IO-bound workloads, it  results in an unpredictable IO performance reported  in several research works conducted on Xen-based virtualized clouds, making it very difficult to predict VMs' IO performance in advance. This is because the Xen scheduler treats all VMs/vCPUs as if  they were equally important,  trying to improve all VMs IO without any discrimination.  (They all compete for better IO). Therefore, LO-crit  vCPUs can adversely impact the IO performance of HI-crit vCPUs if they are co-located. Figure ~\ref{motiv_xen} shows the response times of a Memcached VM when its alone compared to when it runs alongside  CPU-bound VMs. In this experiment, all VMs have enough CPU budgets. As shown, response times are significantly impacted by neighbour VMs,  forcing cloud providers  to resort to over-provisioning to deliver a predictable IO. This is  because neighbor CPU-bound VMs lengthen CALs of the Memcached vCPU and consequently the response times of the Memcached VM. We define CAL as the latency between the time when an I/O request arrives and the time when  its corresponding vCPU is executed on a pCPU.

\textbf{Linux CPU scheduler.} The Completely Fair Scheduler (CFS), leveraged by KVM, is another dominant CPU scheduler in clouds whose main goal is to share processor resources fairly. It treats  vCPUs the same as Linux processes, referring to them as tasks.  The scheduler assigns each task an attribute called vRuntime to track the execution times of tasks. A task's vRurntime inflates as the task runs on a CPU. The speed of inflation depends on the task priority. The higher the priority,  the slower the inflations. The scheduler enforces fairness by keeping the vRuntime of tasks equal. To do so, CFS uses a red-black tree to sort tasks based on their vRuntimes. At each scheduling decision, it chooses the task with the lowest vRuntime (the leftmost task on the tree) to execute next. Since IO bound tasks tend to consume less CPU time relative to CPU-bound ones, their vRuntimes grow slower and therefore they are typically located on the left-most of the red-black tree, being the first to be executed, delivering a good performance for IO-bound workloads. This scheduler does not offer different service qualities for IO-bound tasks given its best-effort nature to enhance IO performance. However, one can  assign different priorities to CPU bound workloads. This scheduler does not offer any schedulability test regarding IO bound workloads. Similar to Xen's CPU scheduler,  it does not support the notion of criticality. Therefore, low criticality tasks impact the QoS of HI-crit tasks. As  shown in Figure ~\ref{motiv_kvm}, under KVM, the quality of Memcached VM as a HI-crit VM is notably impacted by neighbor CPU-bound VMs.

\textbf{Fixed priority schedulers.}  Simplistic fixed-priority  schedulers are not able to offer all requirements outlined in Section 2.1: Fairness, different IO qualities, high utilization and schedulability test.  A promising fixed-priority scheduler is the Rate monotonic (RM) scheduler, which is used in RT-Xen scheduler \cite{RT-Xen}. If RM is used for virtualized clouds, it characterizes vCPUs using a budget (C) and a period (T). Budget C indicates the CPU share that must be allocated to a vCPU during each period T. vCPUs with lower periods get higher priorities. By adjusting C and T, one can assign different quotas of CPU shares to different vCPUs/VMs, offering different execution times to CPU-bound VMs and enforcing fairness. For example,  if  for vCPU v, C = 25 and T = 100, the vCPU will get 25\% of the CPU time. Further,  by adjusting T, the  cloud provider can adjust the CPU access latencies of vCPUs because over each T the vCPU certainly gets access to  a CPU, offering different service qualities to IO-bound workloads, determined by the periods(T) of their corresponding VMs. Finally, RM features a schedulability test that determines if a vCPU set is schedulable or not ($ \sum_{i=1}^{n} \frac{C_i}{T_i} \leq n\times(\sqrt[n]{2} - 1)  $, where n is the number of vCPUs). RM treats all vCPUs equally important,   implying  that  a  vCPU  will  possibly face lengthened CALS if  another  vCPU  fails  to  be  bounded  by  its own  budget. Therefore, in order to deliver a predictable IO performance, system designers have to reserve exaggerated large amounts of CPU time (budgets) for all vCPUs including non-critical vCPUs, which highly exceed the actual worst case execution times to ensure every vCPU performs correctly even under harsh circumstances, ultimately  leading to low utilization.  \cite{Vestal} \cite{Baruah} \cite{EDF}.  

\textbf{Earliest deadline first (EDF)} is a dynamic priority CPU scheduling policy that dynamically assigns highest priority to the tasks with the earliest deadline. Similar to RM, if EDF is used as hypervisor CPU scheduler, it assigns each vCPU an execution time (C) and a period (T) that can be leveraged to offer different QoS to both IO- and CPU-bound workloads. EDF also features a schedulability test $ \sum_{i=1}^{n} \frac{C_i}{T_i} \leq 1$, where n is the number of vCPUs. However, similar to RM, EDF does not consider vCPU criticalities. Therefore, the cloud administrators have to reserve budgets required for peak loads  for all vCPUs to avoid the impacts of vCPUs on HI-crit vCPUs, severe wasting of processor resources that classic scheduling policies such as EDF and RM are not able to mitigate. \cite{Vestal} \cite{EDF}.

\begin{figure}[t]
\centering
  \includegraphics [width= 8cm]{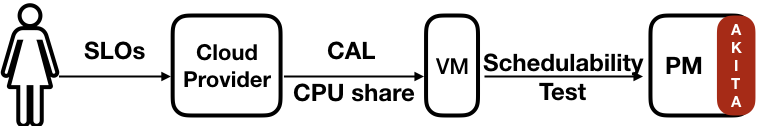}
\caption{ Akita in a virtualized cloud }
\label{dsn}       
\end{figure}
\begin{figure}[t]
\centering
  \includegraphics [width= 7cm]{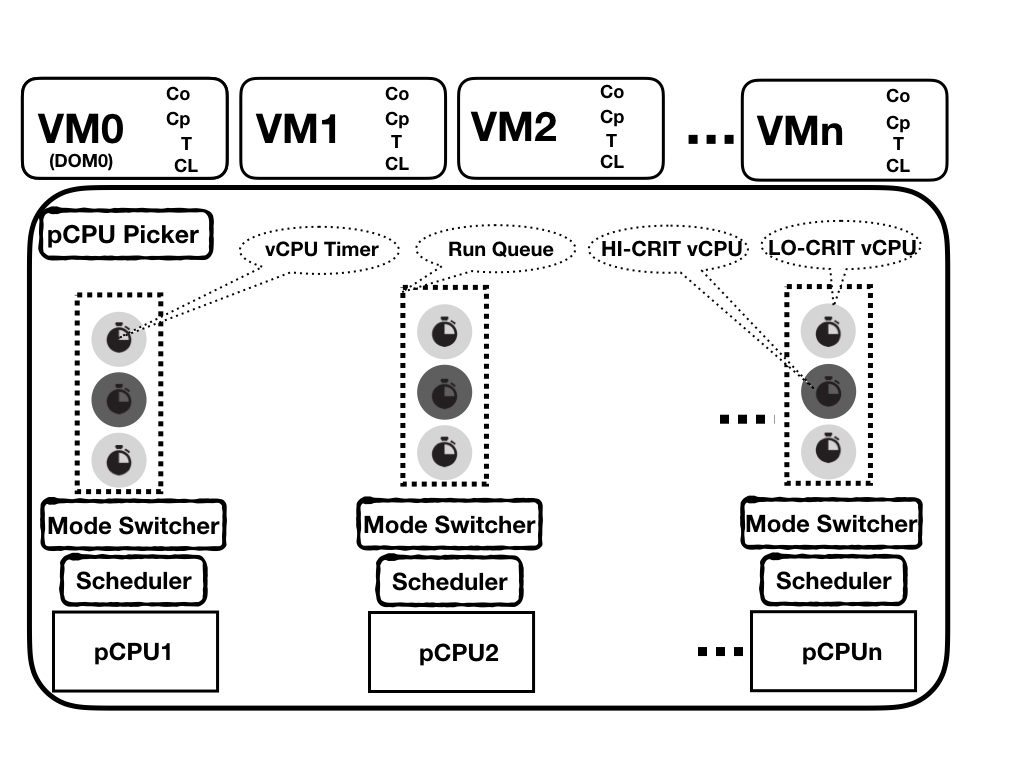}
\caption{ The Architecture  of Akita }
\label{arch}       
\end{figure}

\section{Akita}
Using RM and EDF for vCPU scheduling and strict budget enforcement, a misbehaving VM cannot impact other  VMs. However, if a  VM  does not have enough budget to handle a temporary overload . It  ends  up lagging behind in all future activations. Solving this problem by employing  classic periodic budgeting requires always assigning enough budgets to handle overloads, which is overly pessimistic and leads to resource under-utilization, the main motivation behind designing Akita.

In Akita, we categorize VMs into high-criticality and low-criticality VMs. High-criticality VMs and consequently their vCPUs are characterized by three parameters: an optimistic budget ($C_{opt}$), a pessimistic budget ($C_{pes}$), and a  period ($T$). Low-criticality VMs/vCPUs, on the other hand,  are characterized by an optimistic budget and a period. The optimistic budget of a VM indicates the average CPU time the VM needs, which is less than pessimistic budget that shows the highest CPU time  the high-criticality VM would require under harsh circumstances (peak load).  

Having our schedulable entities (vCPUs) characterized, Akita's scheduling policy is as follows: 

for each vCPU $v$ with period $T$, Akita allocates $v$ its optimistic budget over $T$. Meanwhile, Akita monitors high-criticality vCPUs. If $v$ is a high-criticality vCPU and has consumed its optimistic budget and requires its pessimistic budget, Akita switches to the mode that discards all low-criticality vCPUs to merely schedule high-criticality ones to allocate them their pessimistic budgets. When high-criticality vCPUs no longer need their pessimistic budgets, Akita resets to the mode where all vCPUs are scheduled and allocated their optimistic budgets regardless of their criticalities.

We will discuss the intuition behind Akita's scheduling policy and its schedulability test in Section 2.3. Akita addresses the requirements outlined in Section 2.3 as follows:
 
\textbf{Fairness.} The fraction of CPU time allocated to each high-criticality vCPU is [$C_{opt}/T$, $C_{pes}/T$], depending on the demand of the high-criticality vCPU. The fraction of CPU time allocated to low-criticality vCPUs is [$0$, $C_{opt}/T$], depending on the behavior and existence of high-criticality vCPUs. Note that the CPU shares of high-criticality vCPUs are not impacted by low-criticality vCPUs while the CPU share of low-criticality vCPUs are impacted by high-criticality vCPUs. By adjusting $C$ and $T$ and the criticality level, cloud providers offer  different CPU shares to different VMs. 

\textbf{Different IO qualities.}
In  Akita, each high-criticality vCPU is guaranteed to get access to a CPU over its period, and low criticality vCPUs typically get access to a CPU over their periods, depending on the behavior of high-critically vCPUs.  Therefore, by adjusting the period and the criticality level, cloud providers  offer different IO qualities ranging from predictable to best-effort performance.

\textbf{High utilization.} 
In Akita, both high and low-criticality vCPUs are consolidated on the same CPU and the performance of high-criticality workloads are not impacted by low-criticality vCPUs. This translates to higher utilization while delivering a predictable performance for high-criticality vCPUs, as apposed to existing schedulers that force the cloud providers to isolate high-criticality VMs through over-provisioning in order to achieve predictability.  

\subsection{Akita's Design}
Figure ~\ref{dsn} depicts where Akita stands in a  virtualized cloud. Cloud providers translate VMs' SLOs to  Akita's language, namely pessimistic and optimistic budget, criticality level and period.

Figure ~\ref{arch} illustrates the architecture of Akita. Akita offers an API to cloud providers to specify VMs using four parameters: optimistic budget,  pessimistic budget, period and criticality level. vCPUs inherit these parameters from their corresponding VMs. When a VM is created, the CPU picker assigns each vCPU to a CPU core that can accommodate the vCPU requirements according to Akita's scheduling policy. The design of the CPU picker is described in Section \ref{ssec:pick}. Akita vCPUs periodically ask for CPU shares.   In Section \ref{ssec:pvCPU}, we will explain how we inject periodic behaviour in our vCPUs. Akita's CPUs switch between two modes: normal and high-criticality. In  normal mode, Akita schedules all vCPUs irrespective of their criticalities. In the high-criticality mode, it merely schedules high-criticality vCPUs.   In section \ref{ssec:ms}, we will describe our policy and its corresponding mechanism for CPU mode switching. Finally, we will describe our scheduling algorithm and its schedulability test in section  \ref{ssec:sched}.

\subsection{Periodic Virtual CPUs} \label{ssec:pvCPU}
  Unlike sporadic/periodic  tasks, vCPU do not have a periodic behavior. Their activations are bursty without a clear minimum inter-arrival time.  To inject periodic behavior, we transform over vCPUs to periodic servers that are given a budget and a period.  These budgets are strictly enforced by the scheduler; therefore,  the problem of overrun cannot occur. vCPUs, at the beginning of each period, ask for a CPU share that should be received before the beginning of the next period.  Therefore, Akita imitates  the implicit-deadline sporadic tasks systems in which tasks' deadlines are equal to their periods. 

To enable periodic behavior, Akita's vCPUs feature an internal timer that fires periodically at the beginning of each period. When  a vCPU's timer tick, the vCPU's budget is set to  the optimistic budget If the vCPU is a LO-crit vCPU; otherwise, the vCPU is a HI-crit vCPU whose budget is set to its pessimistic budget. 

As a vCPU runs on a pCPU, its budget is decreased proportionally. If a vCPU's budget runs out. The vCPU is deactivated, meaning that the scheduler ignores the vCPU if there exist other vCPUs with positive budgets. At the next tick of the vCPU's timer, the vCPU is activated again and its budget is set to  its optimistic  or  pessimistic budget, depending on its criticality level. 

Akita's scheduler is work conserving. When there is no  active vCPU, It schedules inactive vCPUs, suggesting that Akita will not remain idle if there exist runnable vCPUs.  

\subsection{Mode Switching} \label{ssec:ms}
Each pCPU starts in the LO-crit mode. As a consequence, the scheduler at first treats all vCPUs as equally important and allocates each vCPU its desired optimistic CPU budget. Meanwhile, using our accounting mechanism, we monitor the CPU consumption of the currently running vCPU. If the currently running vCPU is a HI-crit vCPU that has been executed for its optimistic  budget and is still runnable (there are running$\&$ready processes inside its corresponding VM), it  indicates  the HI-crit vCPU requires its  pessimistic  CPU share. If so, a mode switch is immediately triggered and the scheduler switches  to the HI-crit mode. Henceforth, the pCPU behaves as if there were only HI-crit vCPUs; It merely schedules them to allocate them their pessimistic budgets and   LO-crit vCPUs will not receive any further execution.

In the HI-crit mode, an early mode switch to  the LO-crit mode may impact the performance of HI-crit vCPUs. Therefore, before switching back to the  LO-crit mode, we need to make sure that HI-crit vCPUs   no longer need their  pessimistic budgets. To this end, we assign each HI-crit vCPU a temperature. When a HI-crit vCPU needs its pessimistic budget (has received its optimistic budget and does not signal completion), its temperature is set to a positive number.  On the other hand, if a HI-crit vCPU whose temperature is greater than zero has not asked for its pessimistic budget in the last period, we decrease the vCPU's temperature by one degree, cooling down the vCPU. The temperature of a HI-crit vCPU, in fact, indicates how recently the vCPU has asked for its HI-crit CPU share. The hotter a vCPU, the more recent it caused a mode switch to the HI-crit mode.

 Having known vCPU temperatures, our approach for switching back to the LO-crit mode is straightforward. If there exists no vCPU  with positive temperature, it indicates that recently no vCPU has requested for its pessimistic CPU share  and therefore the pCPU's mode can be switched back to the LO-crit mode.  The  initial value of the temperature determines how fast a pCPU gets reset to LO-crit mode. The higher the  temperature, the slower switching back to the LO-crit mode.   Therefore, in Akita, switching to the HI-crit mode is instant while switching back to the  LO-crit mode is conservative, prioritizing the predictability and performance of HI-crit vCPUs and VMs.  
 \subsection{Scheduling } \label{ssec:sched}
The main objective of the scheduling unit is to choose the next vCPU from a pCPU's run queue to execute for a predefined amount of time known as the time slice. In fact, the scheduling function determines the order of scheduling of vCPUs located on the run queue. In Akita, the scheduling  function is invoked for the following reasons:  

(1) When  a vCPU's timer ticks, the vCPU becomes activated; since it is likely that the newly activated vCPU is more important/urgent than the currently running vCPU, the scheduler is invoked (2)  The currently running vCPU relinquishes the pCPU voluntarily for a reason (e.g., the vCPU is idle), the scheduler is invoked to choose the next runnable vCPU (3)  The currently running vCPU relinquishes the pCPU forcibly because it has used its budget (4)  and finally when a  HI-crit vCPU has received its optimistic budget, the scheduler is invoked to check if the vCPU is still runnable; if so, a mode  switch   must be triggered.

At each scheduling decision Akita takes several steps. First, it  determines the pCPU's mode. Second, It  updates the budget of the currently running vCPU based on its CPU consumption. Third,  it updates the state (active when the vCPU's budget exhausts; otherwise inactive) and the temperature of the currently running vCPU.  Finally, It  inserts the currently running vCPU into the run-queue to choose the  next eligible vCPU  from the run queue. If mode is HI-crit, Akita chooses an active HI-crit vCPU with the earliest deadline, if mode is LO-crit, it chooses the vCPU with earliest deadline regardless of its criticality. The time slice is set to the budget of the chosen vCPU. vCPUs deadlines are set to their periods when their timer tick. Therefore, at each mode, Akita imitates EDF scheduling of implicit deadline sporadic task systems. 

vCPUs are sorted based on their deadlines, and active vCPUs are always located before inactive ones. The time complexity of vCPU insertion is $O_{(n)}$. At each scheduling decision, the scheduler  picks the vCPU at the head of runqueue to execute next with time complexity of $O_{(1)}$. Therefore, similar to Xen's Credit schedulers Akita is an $O_{(n)}$ scheduler. 

\subsection{ Akita's schedulability  test} \label{ssec:test}

Low utilization caused by traditional real-time policies such as EDF and RM  scheduling  led researchers to design  a novel scheduling policy, namely mixed-criticality scheduling (MCS) which is the main  intuition behind Akita's design. MCS discriminates between important and non-important tasks by assigning a new dimension to tasks, referred to as criticality.
 MC scheduler goal is to allocate each task   its optimistic budget over each period until a HI-crit task asks for its pessimistic budget. If So, the scheduler will move to   HI-crit mode where all LO-crit tasks are   immediately discarded.  An MC scheduler not only must favor urgent jobs (e.g., jobs with lower periods in RM) but also prioritize high-critical jobs to prepare them for potentially long executions. The compromise between urgency and importance results in exponential choices, leading to the fact that mixed-criticality scheduling is NP-hard in a strong sense \cite{MCS} \cite{practical}.  Therefore, several approximation algorithms have been proposed for scheduling of MC tasks including but not limited to EDF-VD \cite{EDF} and OCBP  \cite{OCBP}.

EDF-VD   leverages  EDF to schedule tasks. However, it shrinks the deadlines of HI-crit tasks  by a certain factor so that HI-crit tasks will be promoted by the EDF scheduler. In fact, it reduces the actual deadlines (D) of HI-crit task jobs to modified deadlines that are called virtual deadlines (VD) which are lower than actual deadlines. EDF-VD calculates virtual deadlines using Equations (1)-(5), wherein $C_{opt}$ and $C_{pes}$  indicate  optimistic  and  pessimistic budgets, respectively,  $U_1(1) $ is the utilizations of the LO-crit tasks,  $U_2(1)$ is the utilizations of the HI-crit tasks considering their optimistic budgets,  $U_2(2)$ is the utilizations of the HI-crit tasks considering their pessimistic budgets, and $VD$ indicates the virtual deadline calculated by shrinking the actual deadline by a factor of $x$. Condition (6) is the  EDF-VD's schedulability test. If  condition (6) holds for a  task set, the task set is MC-schedulable. The key property of real-time scheduling strategies is that they guarantee that the deadlines of all the tasks are met. This requirement is too strong for the virtualized clouds, where  infrequent violations of temporal guarantees would not lead to catastrophic consequences.  Although  the current version of Akita  uses EDF-VD to determine the deadline of our vCPUs as well as  our schedulability test, a cloud provider can take a less conservative schedulability test (e.g., EDF schedulability test). Unlike  MC systems, Akita do not discard LO-crit vCPUs when a mode switch happens. Our vCPUs alternate between low and high criticality modes. 

\begin{equation}
U_1(1) =   \sum_{\tau_i:X_i=1}C_{opt}(i)/T_i 
\label{eq:1}
\end{equation}
\begin{equation}
 U_2(1)  =   \sum_{\tau_i:X_i=2}C_{opt}(i)/T_i 
 \label{eq:2}
\end{equation}
\begin{equation}
 U_2(2) =     \sum_{\tau_i:X_i=2}C_{pes}(i)/T_i 
 \label{eq:3}
 \end{equation}
 \begin{equation}
 x = U_2(1) / (1- U_1(1))
 \label{eq:4}
  \end{equation}
 \begin{equation}
{VD} = \begin{cases} now+x * T_i \quad  if &\text{Xi = 2}\\ now + T_i    \quad  \, \,  \quad if &\text{Xi = 1}\\ \end{cases}
 \label{eq:5}
  \end{equation}
  \begin{equation}
  x * U_1(1)+U_2(2) \leq  1
 \label{eq:6}
 \end{equation}

%
%
%
%
%
%
%

\subsection{vCPU Assigning}  \label{ssec:pick}
When a VM is initially consolidated in a PM, Akita invokes  the pCPU picker function to find    appropriate pCPUs that can accommodate  vCPUs of the  new VM. The pCPU picker   uses   our schedulability test to determine if a pCPU can accommodate a vCPU. For each vCPU, our CPU picker  assigns the vCPU to the first CPU core that can accommodate the vCPU.  If the pCPU picker fails to find an appropriate pCPU for   a vCPU, Akita simply notifies the cloud provider, implying  that hosting this VM would lead to SLO violations. Akita, therefore, offers cloud   providers  a  schedulability test that enables  a wiser VM placement, and thus mitigating the need for costly monitoring services . 
More importantly,  Akita's first-fit mechanism for vCPU assigning keeps the number of operating cores as low as possible. The remaining idle cores leverage C-state mechanism to turn off their internal components to save power. Akita's counterparts, on the other hand,    operate all existing CPU cores in a blind effort to increase IO performance. Therefore, adopting Akita will lead to lower power and thus operational costs of virtualized clouds.   

\section{Evaluation}
We evaluate the Akita's performance by answering the following questions quantitatively:
(1)  Is Akita able to offer different service qualities to both IO- and CPU- bound workloads? (~\autoref{ssec:diff})
(2) Do IO-bound HI-crit VMs (e.g., Memcached server VMs) deliver a predictable performance even when they are collocated with other CPU- or latency-bound VMs? (~\autoref{ssec:latency}) 
(3) How does Akita perform when multiple HI-crit VMs are collocated on the same pCPU?  (~\autoref{ssec:latency})
(4) Does Akita keep the CPU shares of HI-crit VMs intact when they are collocated with LO-crit VMs? (~\autoref{ssec:cpu}) 
(5) How does Akita's first-fit mechanism for vCPU assigning perform compared to Linux and Xen process load-balancers?  (~\autoref{ssec:smp})
(6) Does RQS mechanism raise QoS of LO-crit VMs?(~\autoref{ssec:smp})

To answer these questions, we use an experimental testbed that consists of a physical client machine and a physical server machine. The  server and client machines both have 12-core 3.2 GHz  Intel Xeon CPU and 32 GB of memory. The physical server machine hosts guest VMs, and the physical client machine generates workloads for IO-bound VMs. We always monitor the load-generator machine and the network bandwidth to make sure they are not performance bottlenecks. Physical and virtual machines run Linux kernel 3.6. We dedicate two CPU cores to Xen's driver domain (Dom0) in all experiments.  For Akita's experiments, all VMs are configured by four parameters: criticality level, period, optimistic and pessimistic budgets. Note that the LO-crit VMs do not have pessimistic budgets.  In these experiments,  we use Sysbench as  a  CPU-bound benchmark that computes prime numbers. For IO-bound workloads, we interchangeably use Memcached that follows the event-driven concurrency model and Thrift RPC  with the thread-driven concurrency model. Unless otherwise stated, Table  \ref{t_mem} presents the configurations of all VM types used to evaluate Akita. 
\begin{table}[]
\centering
\caption{The configurations of VMs}
\label{t_mem}
\begin{tabular}{  |p{1cm}|p{1.3cm}|p{1.3cm}|p{1.2cm}|p{0.7cm}|p{0.4cm}|}
\hline
\textbf{VMType}                                 &\textbf{$C_{opt}$}       &\textbf{$C_{pes}$}              &\textbf{Period}                         &\textbf{CL}           \\ \hline \hline
Type1                                   &5ms                      &25ms                 &100ms                         &High                  \\ \hline
Type2                                  &25ms                      &-                 &100ms                         &Low                  \\ \hline
Type3                                   &50ms                      &-                 &100ms                         &Low                 \\ \hline
Type4                                   &5ms                      &-                 &50ms                         &Low               \\ \hline
Type5                                   &10ms                      &-                 &100ms                         &Low               \\ \hline
Type6                                   &10ms                      &50                &100ms                         &High               \\ \hline
\end{tabular}
\end{table}

\begin{figure}
\centering
\begin{subfigure}{0.40\linewidth}
\centering
  \includegraphics[width= 3.8cm,height=3.8cm, keepaspectratio]{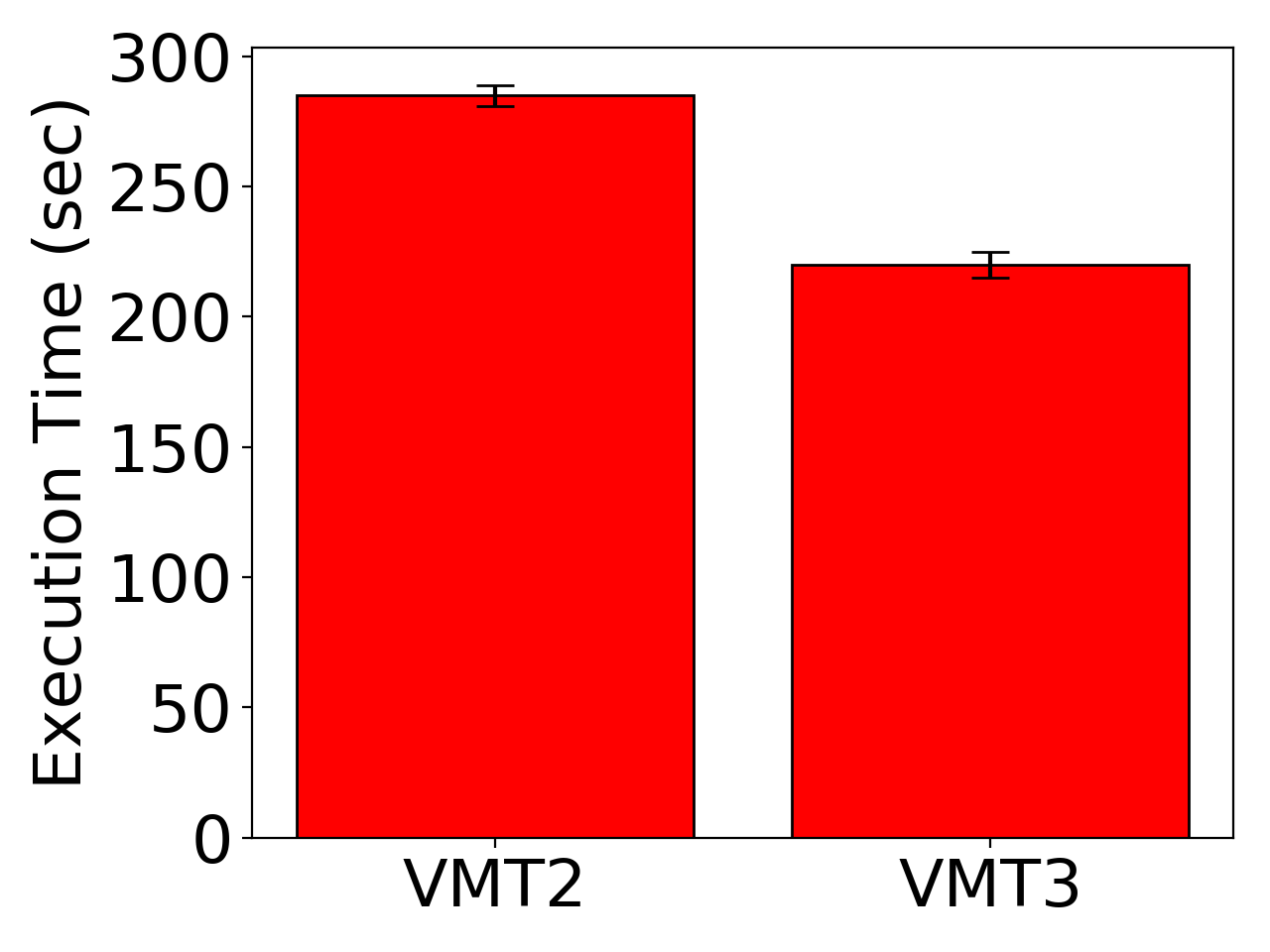}
  \caption{ Execution time}
  \label{diff_CPU}
 \end{subfigure}
 ~
\begin{subfigure}{0.40\linewidth }
\centering
  \includegraphics[width=3.8cm,height=3.8cm, keepaspectratio]{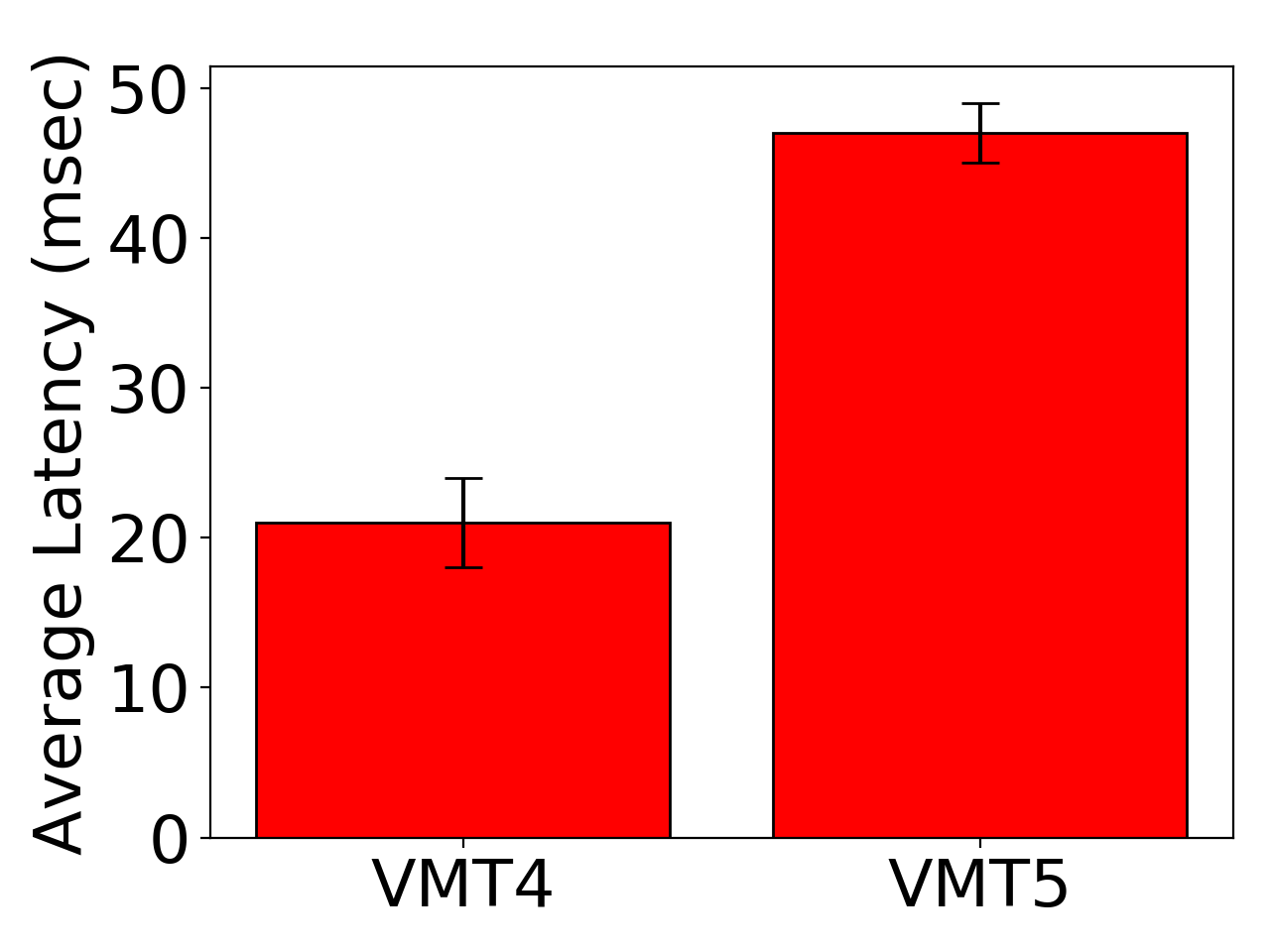}
  \caption{Average of RTTs }
  \label{diff_IO}
\end{subfigure}
\caption{ Differentiated service qualities offered by Akita}
\label{diff}
\end{figure}

\subsection {Differentiated Service Qualities} \label{ssec:diff}
Akita offers different levels of QoS to both IO- and CPU-bound workloads by enforcing different CPU shares adjusted by periods  and  budgets, and different CALs  adjusted by periods. In this experiment, the physical server machine hosts a bunch of CPU-bound (1x Type2 VM and 1x Type3 VM) and IO-bound VMs (1x Type4 VM and 1x Type5 VM). CPU bound VMs run Sysbench, and IO-bound VMs run an RPC benchmark. Each IO-bound VM is stressed with multiple concurrent client threads hosted in the client physical machine, each generating 500 calls/second forming a Poisson process. Figure \ref{diff_CPU} shows the execution times of Sysbench benchmarks, and Figure   \ref{diff_IO} shows the average latency of RPC requests sent to each IO-bound VM. Sysbench benchmark hosted in VMs with higher $budget/period$ have a lower execution time, and RPC workloads hosted in VMs with lower periods (CALs) are more responsive, suggesting that Akita is able to offer different levels of QoS to both IO- and CPU-bound workloads.  

\begin{figure*}
\centering
\begin{subfigure}{0.9\linewidth}
  \centering
  \includegraphics[width= 15cm,height=0.4cm, keepaspectratio]{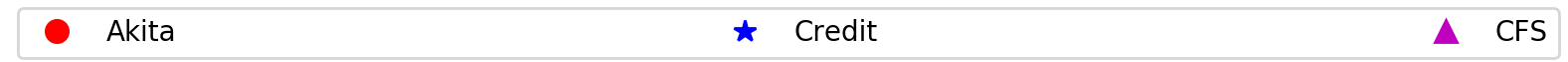}
 \end{subfigure}
\begin{subfigure}{0.23\linewidth}
  \centering
  \includegraphics[width= 4.4cm,height=4.5cm, keepaspectratio]{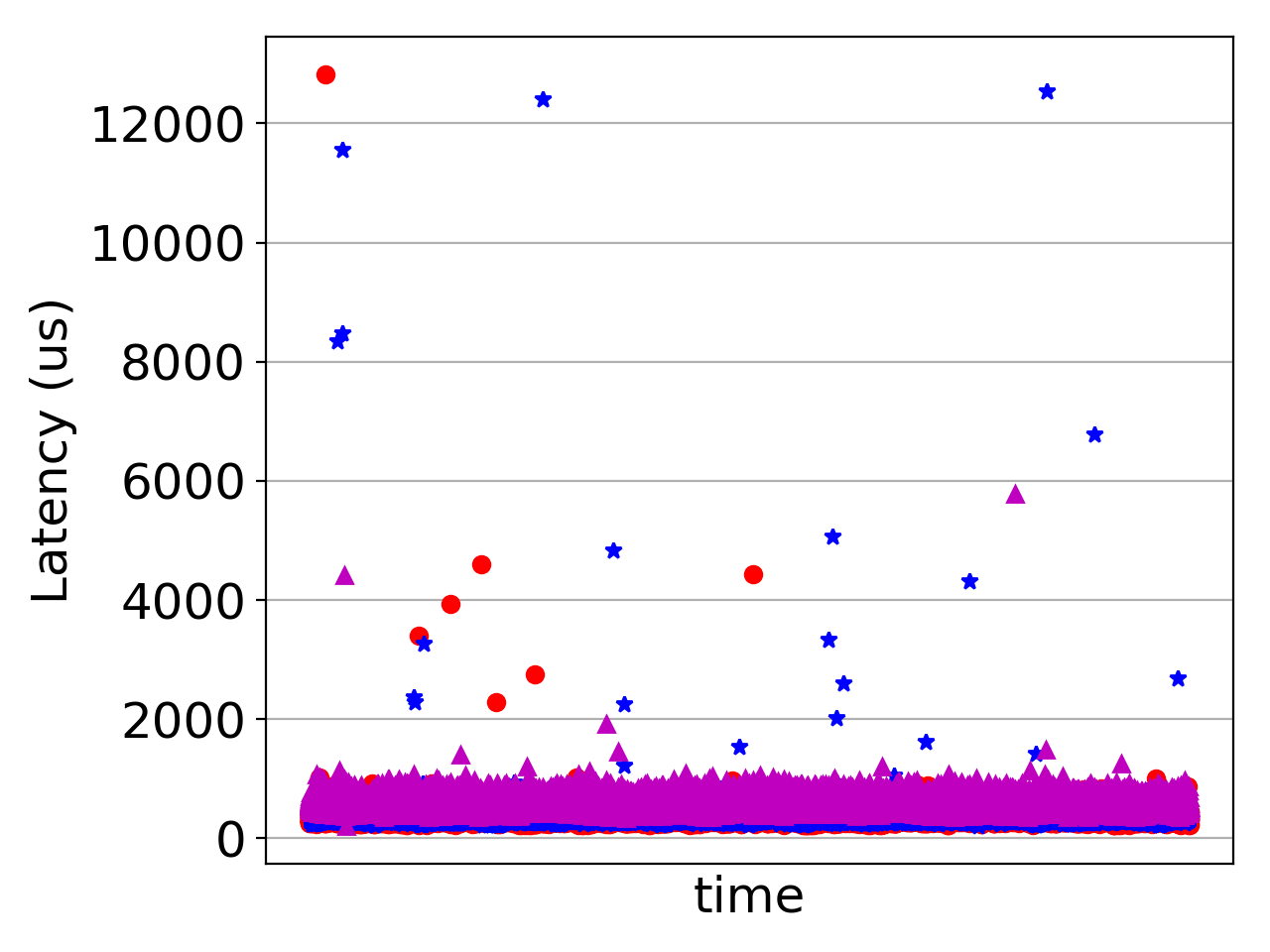}
  \caption{RTTs (Alone)}
  \label{mem_Akita_a}
  ~
\end{subfigure}
\begin{subfigure}{0.23\linewidth }
  \centering
  \includegraphics[width=4.4cm,height=4.4cm, keepaspectratio]{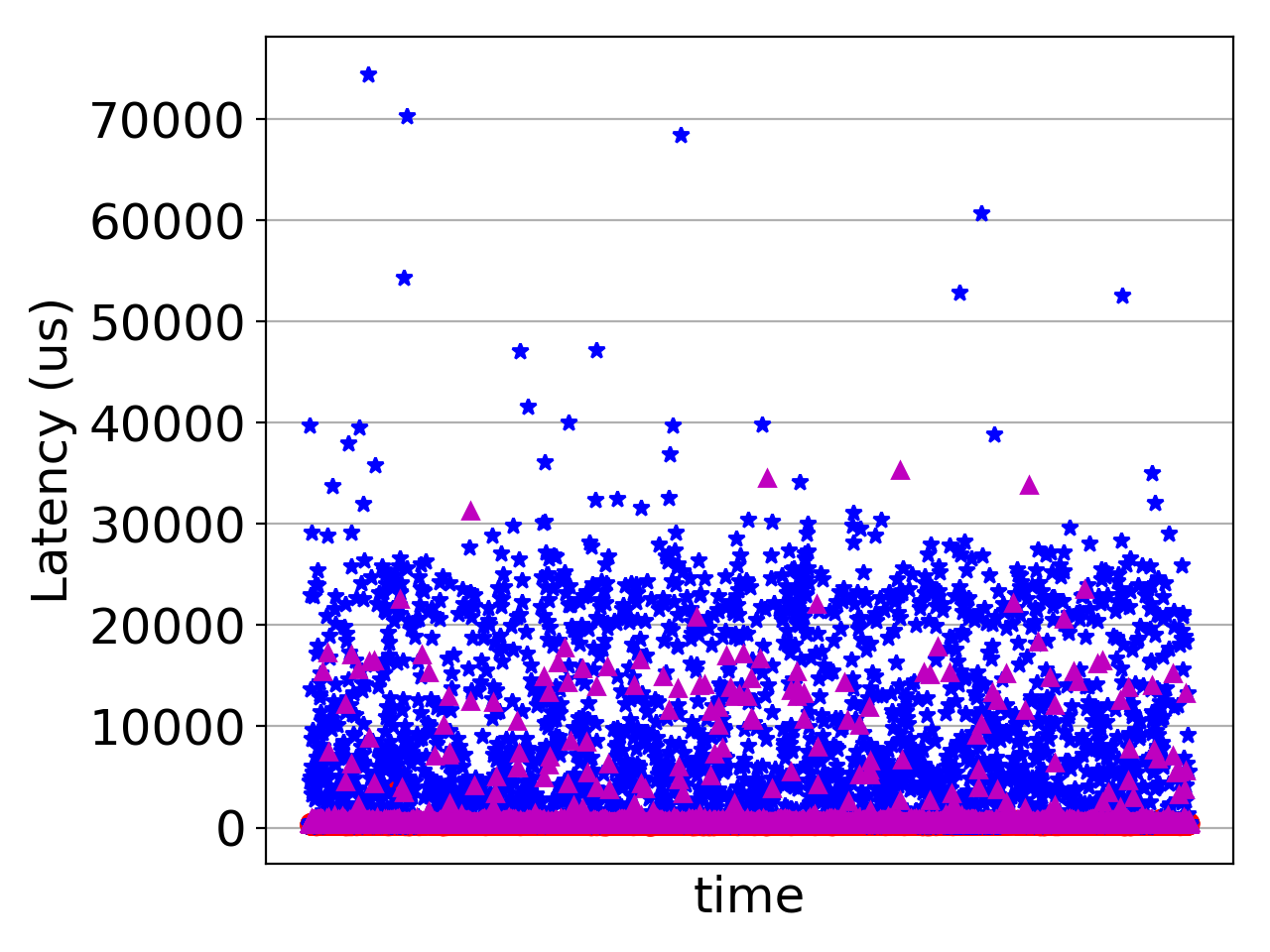}
  \caption{RTTs (Not Alone) }
  \label{mem_Akita_b}
\end{subfigure}
~
\begin{subfigure}{0.23\linewidth }
  \centering
  \includegraphics[width= 4.4cm,height=4.4cm, keepaspectratio]{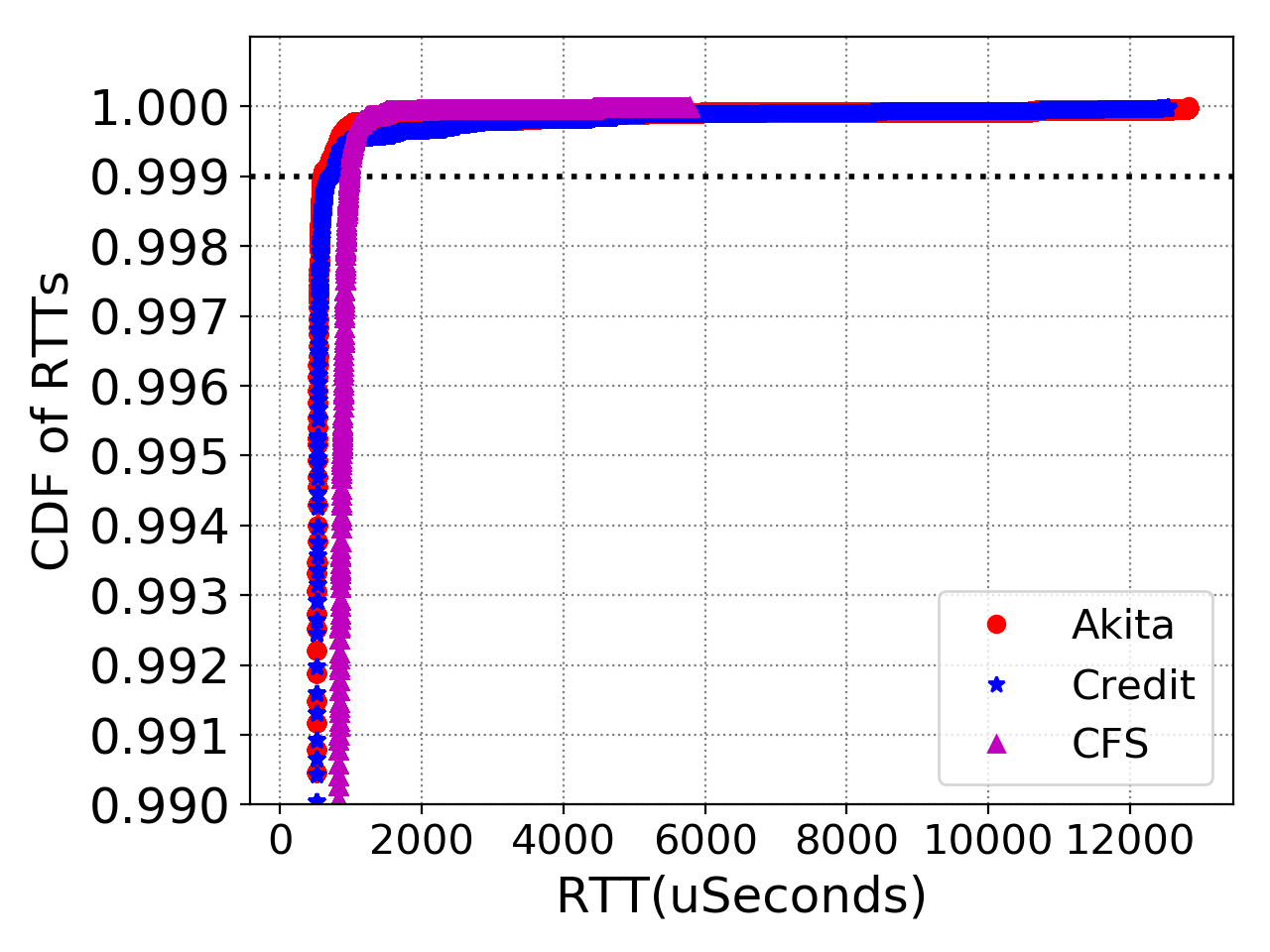}
  \caption{Tail (Alone)}
  \label{mem_Akita_c}
\end{subfigure}
~
\begin{subfigure}{0.23\linewidth}
  \centering
  \includegraphics[width= 4.4cm,height=4.4cm, keepaspectratio]{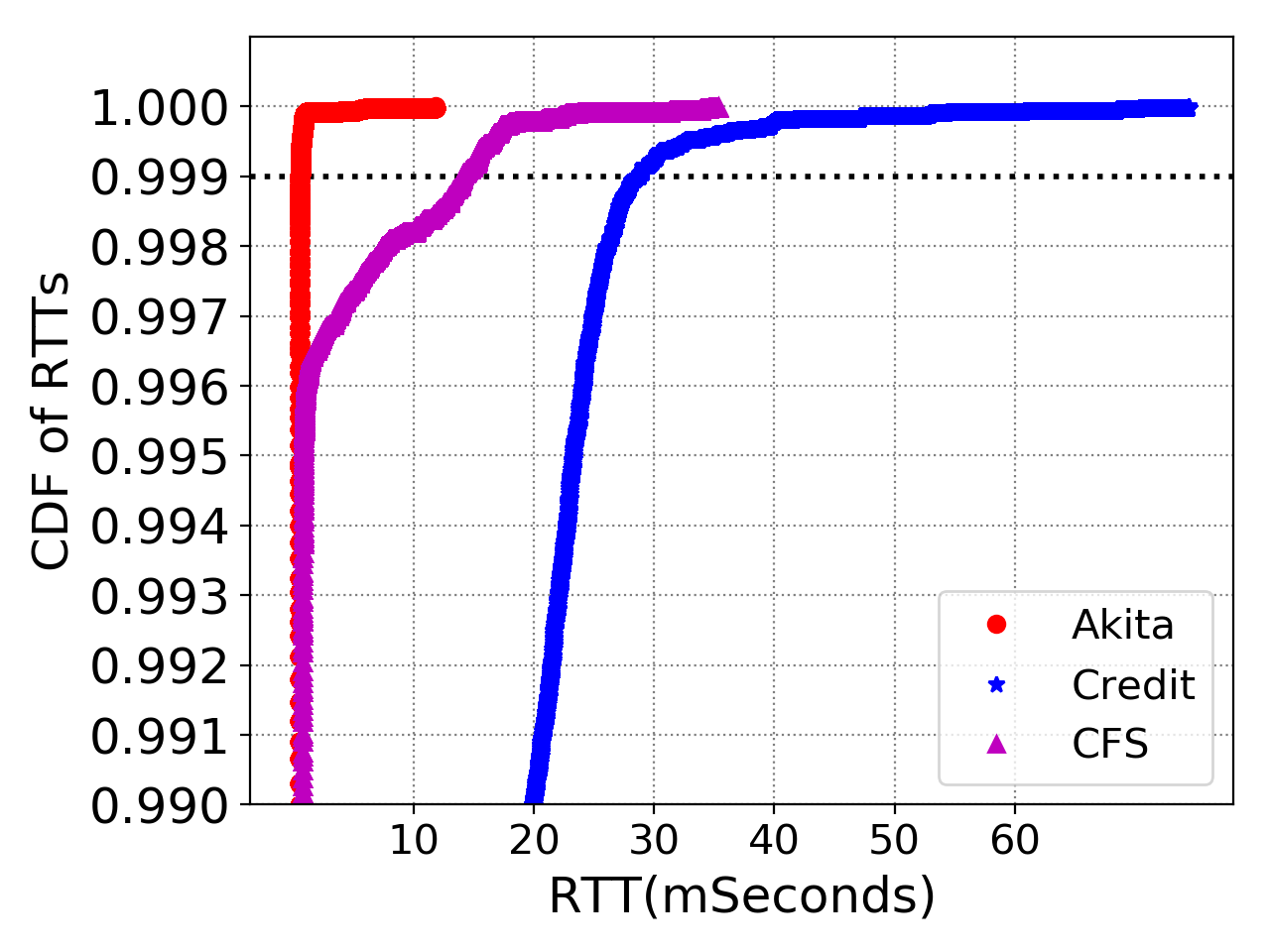}
  \caption{Tail (Not Alone)}
  \label{mem_Akita_d}
\end{subfigure}
\caption{ Performance of a Memcached VM }
\label{mem_Akita}
\end{figure*}

\begin{figure*}
  \centering
\begin{subfigure}{0.9\linewidth}
  \centering
  \includegraphics[width= 15cm,height=0.4cm, keepaspectratio]{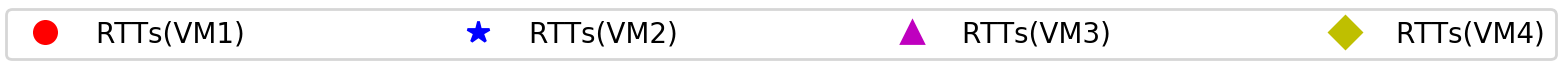}
 \end{subfigure}
\begin{subfigure}{0.22\linewidth}
  \centering
  \includegraphics[width= 4.4cm,height=4.4cm, keepaspectratio]{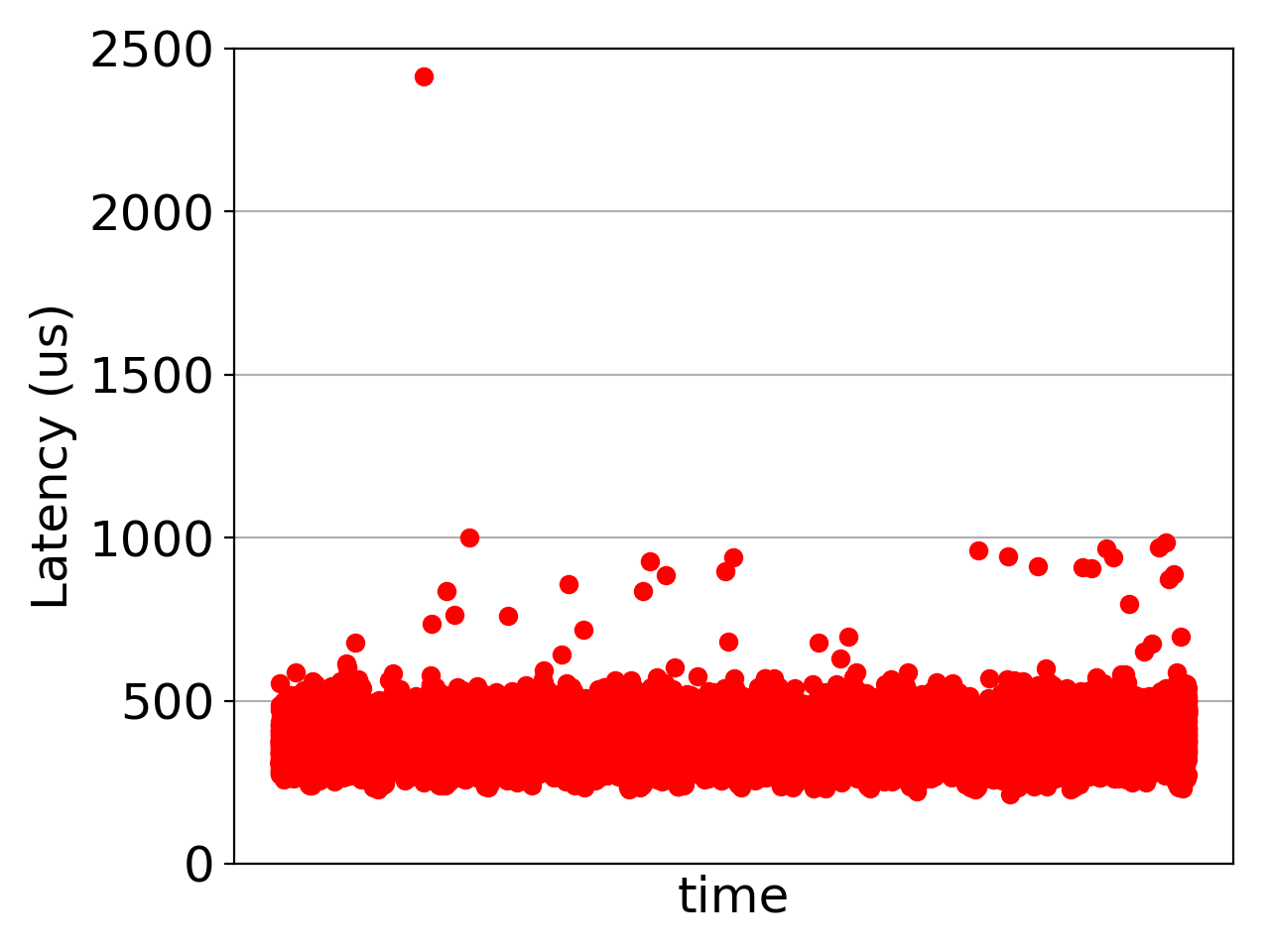}
  \caption{RTTs}
  \label{mem_multi_a}
 \end{subfigure}
 ~
\begin{subfigure}{0.22\linewidth }
  \centering
  \includegraphics[width=4.4cm,height=4.4cm, keepaspectratio]{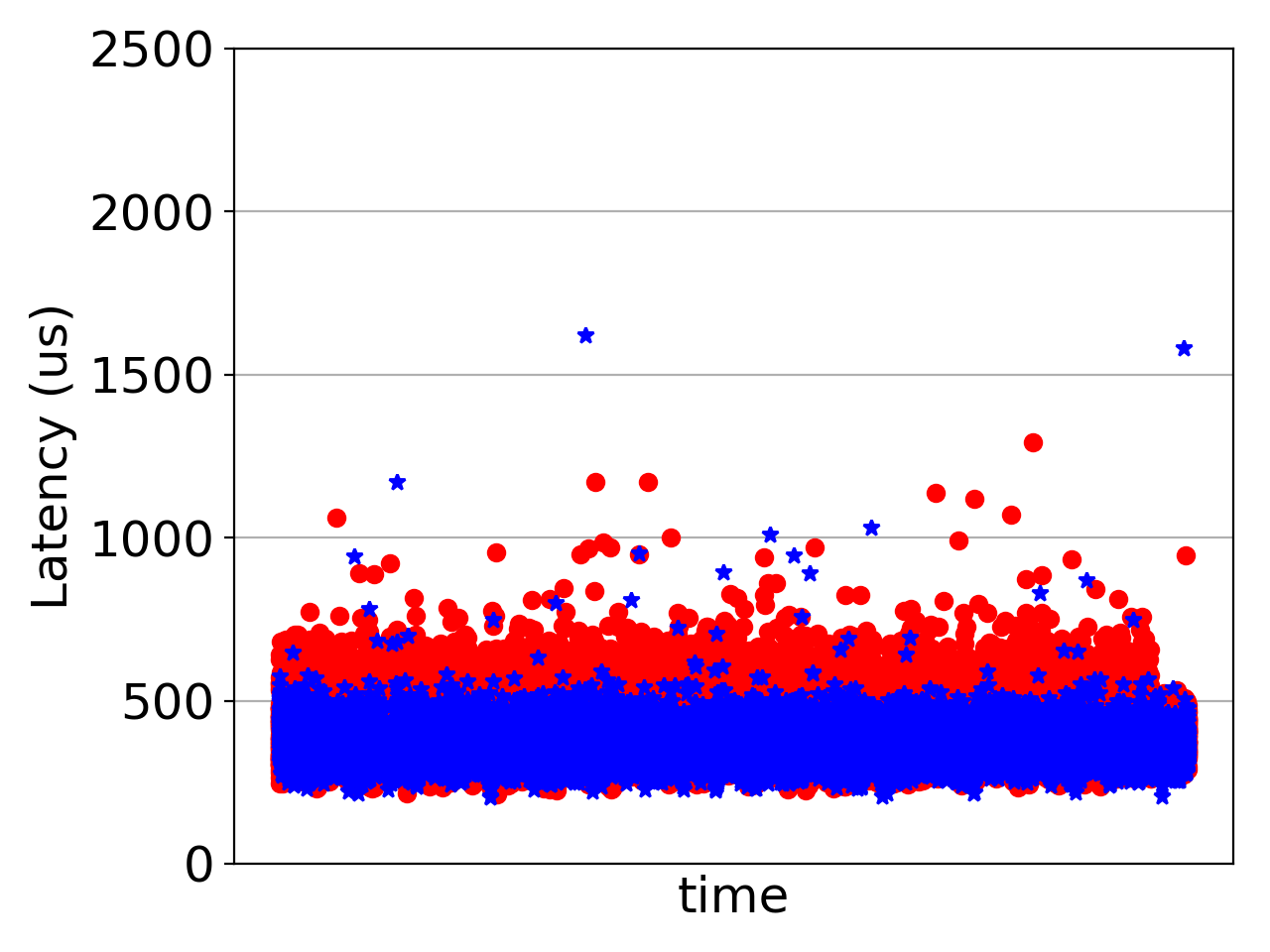}
  \caption{RTTs}
  \label{mem_multi_b}
\end{subfigure}
~
\begin{subfigure}{0.22\linewidth }
  \centering
  \includegraphics[width= 4.4cm,height=4.4cm, keepaspectratio]{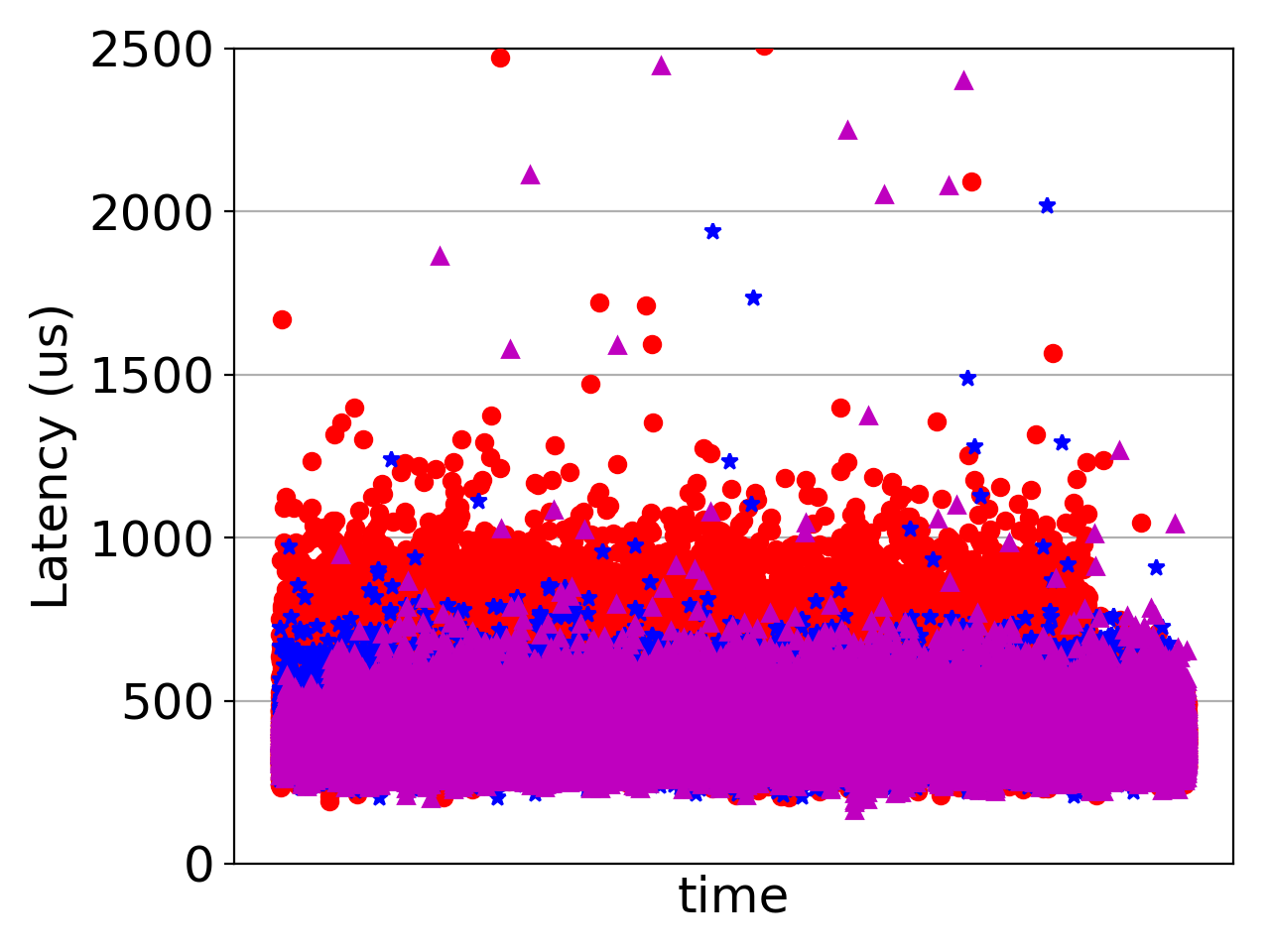}
  \caption{RTTs }
  \label{fmem_multi_c}
\end{subfigure}
~
\begin{subfigure}{0.22\linewidth}
  \centering
  \includegraphics[width= 4.4cm,height=4.4cm, keepaspectratio]{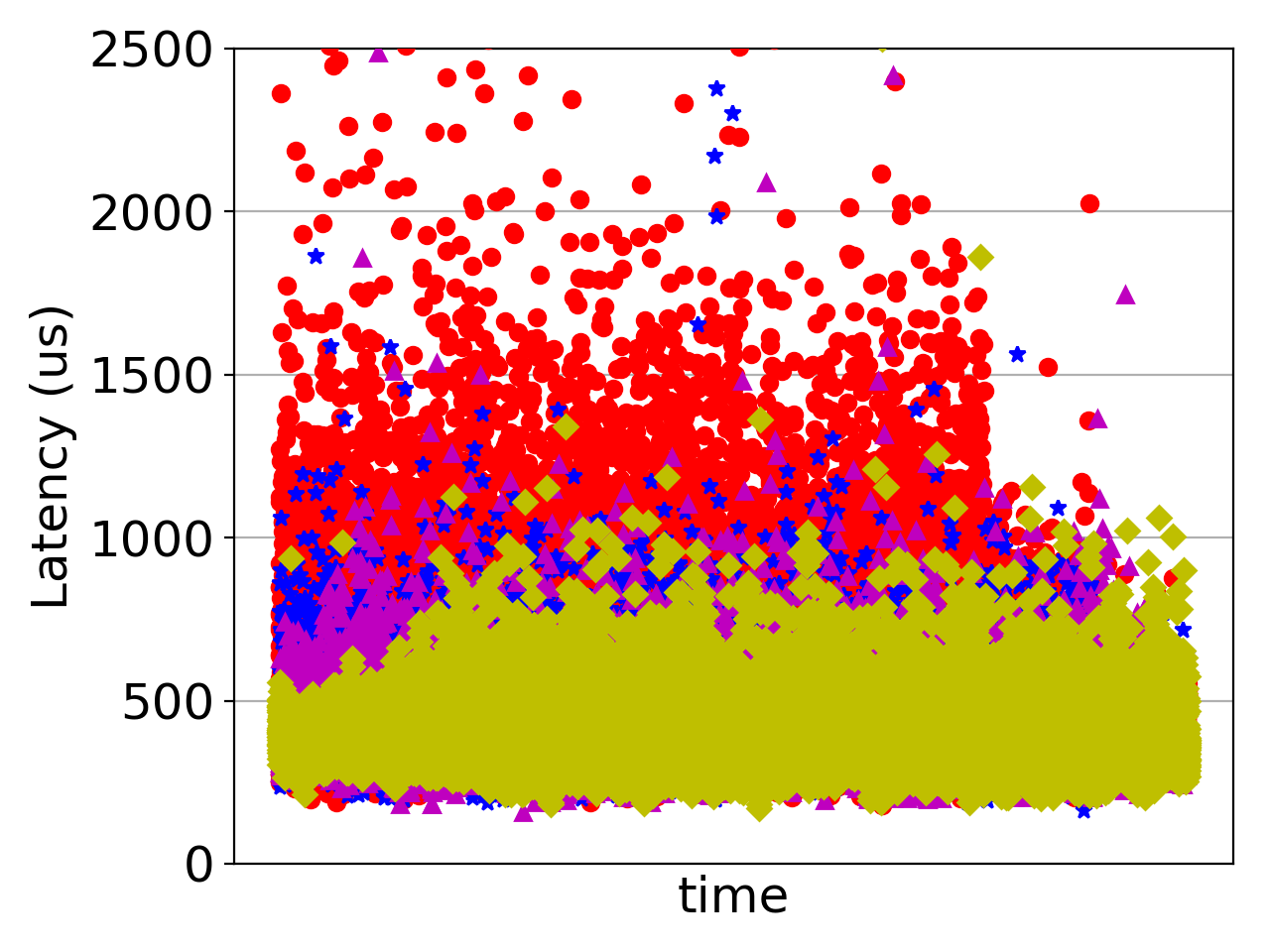}
  \caption{RTTs}
  \label{mem_multi_d}
\end{subfigure}
\caption{ Performance of Akita  when multiple HI-crit  VMs co-execute }
\label{mem_multi}
\end{figure*}
\subsection {HI-crit  IO-bound VMs} \label{ssec:latency}
To assess the delivered QoS of HI-crit IO-bound VMs under Akita, we use a Memcached server VM (Type1). This VM is stressed for two minutes by 50 concurrent connections, each sending 700 requests per second with exponential inter-arrivals. Figure \ref{mem_Akita_a} shows the RTTs of the Memcached server VM when it is running alone on a pCPU under Akita, Credit, and CFS (for Credit and CFS experiments, we use a VM with the default configuration). As expected, RTTs are predictable and rarely higher  than 5ms under all schedulers. This is because the Memcached vCPU runs alone on a pCPU, meaning that its CALs remain intact which results in predictable and fast response times.  In this experiment, the Memcached VM utilizes around 25\% of the CPU time to accommodate the incoming key-value workload.  

Next, we host three CPU-bound VMs (3x Type2 VMs) running Lookbusy- a synthetic CPU-intensive workload alongside the Memcached VM. Each of these VMs utilizes around 25\% of the pCPU time during the experiment, 75\% combined.  For the Akita experiment, we configure the Memcached VM as a HI-crit VM and the CPU-bound VMs as LO-crit VMs. For CFS and Credit experiments, we use the default configuration for all VMs. Given the configuration of these VMs (see Table  \ref{t_mem}), Akita packs all of these VMs on same CPU core based on its first-fit assigning mechanism. To allow for a fair comparison, we pin all VMs on the same CPU core under Credit and CFS.    We then stress the Memcached server VM as before, and record RTTs of Memcached requests under all schedulers. 

Figure \ref{mem_Akita_b} shows the RTTs of the Memcached server VM. Under Akita, the Memcached server VM still delivers  predictable performance (response time) even though it is collocated with three other LO-crit VMs, suggesting that under Akita, the expected QoS of HI-crit latency-bound VMs is not influenced by collocated LO-crit VMs whereas the RTTs of Memcached requests are  deteriorated by collocated VMs under both Xen and KVM. Figure \ref{mem_Akita_c} and   \ref{mem_Akita_d} show the last latency percentiles of Memcached server VM when it is alone on a pCPU compared to when it runs alongside CPU-bound VMs. Under Akita, the tail latency at 99.9 percentile remains intact as if the Memcached server VM ran alone on the pCPU while the tail latency is increased by 96\% and 93\% under Xen and KVM, respectively. This is because Akita keeps the CALs of Memcached vCPU as a HI-crit vCPU under the expected value (period), as it switches the pCPU's mode to HI-crit mode  where LO-crit vCPUs are ignored to make sure the HI-crit vCPU gets its desired CPU share, as well as geting access to the pCPU before the deadline specified by its period.   Under best-effort scheduling policies of Xen and KVM, on the other hand, collocated CPU-bound vCPUs impose lengthened CALs, resulting in variable performance of the Memcached VM.

We next examine the effectiveness of Akita when multiple HI-crit IO-bound VMs run on a pCPU. At first, a Memcached server VM (Type2) which is tagged as a LO-crit VM is stressed by four concurrent connections, each sending 600 requests per second. We then  host another Memcached server VM (Type1) that is tagged as a HI-crit VM and stress both of the Memcached VMs using 8 concurrent threads. Figure \ref{mem_multi_b} shows the RTTs of Memcached VMs. The RTTs of the LO-crit VM slightly fluctuate, while the HI-crit Memcached VM delivers a predictable (expected) performance. We increase the number of HI-crit Memcached VMs from 1 VM up to 3 VMs (3x Type1 VMs) and repeat the experiment. As the number of HI-crit VMs grows, the RTTs of the LO-crit VMs swing more noticeably. However, in our experiment, no VM misses deadlines because the RTTs of IO-bounds VMs mainly depend on CALs of their corresponding vCPUs. When all vCPUs are latency-sensitive, they remain on a pCPU for a very short time ($\sim$46 $\mu s$) and then voluntarily relinquish the pCPU. They, therefore, cannot result in long CALs for one another, the reason why no VM misses deadlines in this experiment, as depicted in Figure  \ref{mem_multi}. 
\begin{figure}[h]
  \centering
	\includegraphics[width= 6cm,height=3cm]{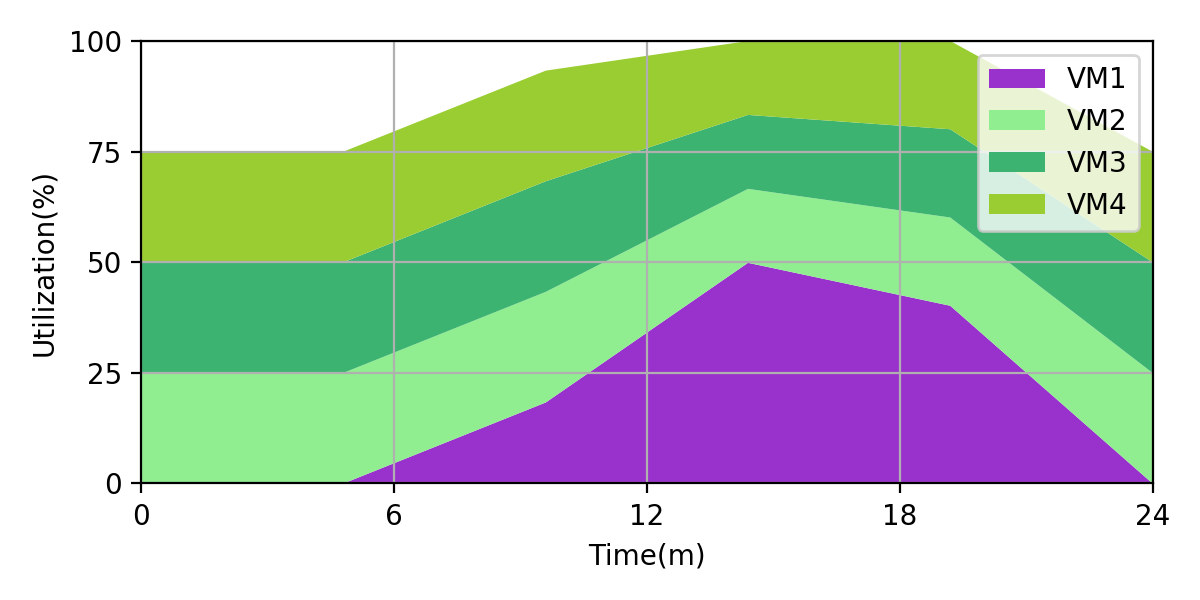}
       \caption{CPU utilization of VMs} 
	\label{cpu}
\end{figure}

\begin{figure*}
\begin{subfigure}{0.30\linewidth}
  \centering
  \includegraphics[width= 5cm,height=4.5cm, keepaspectratio]{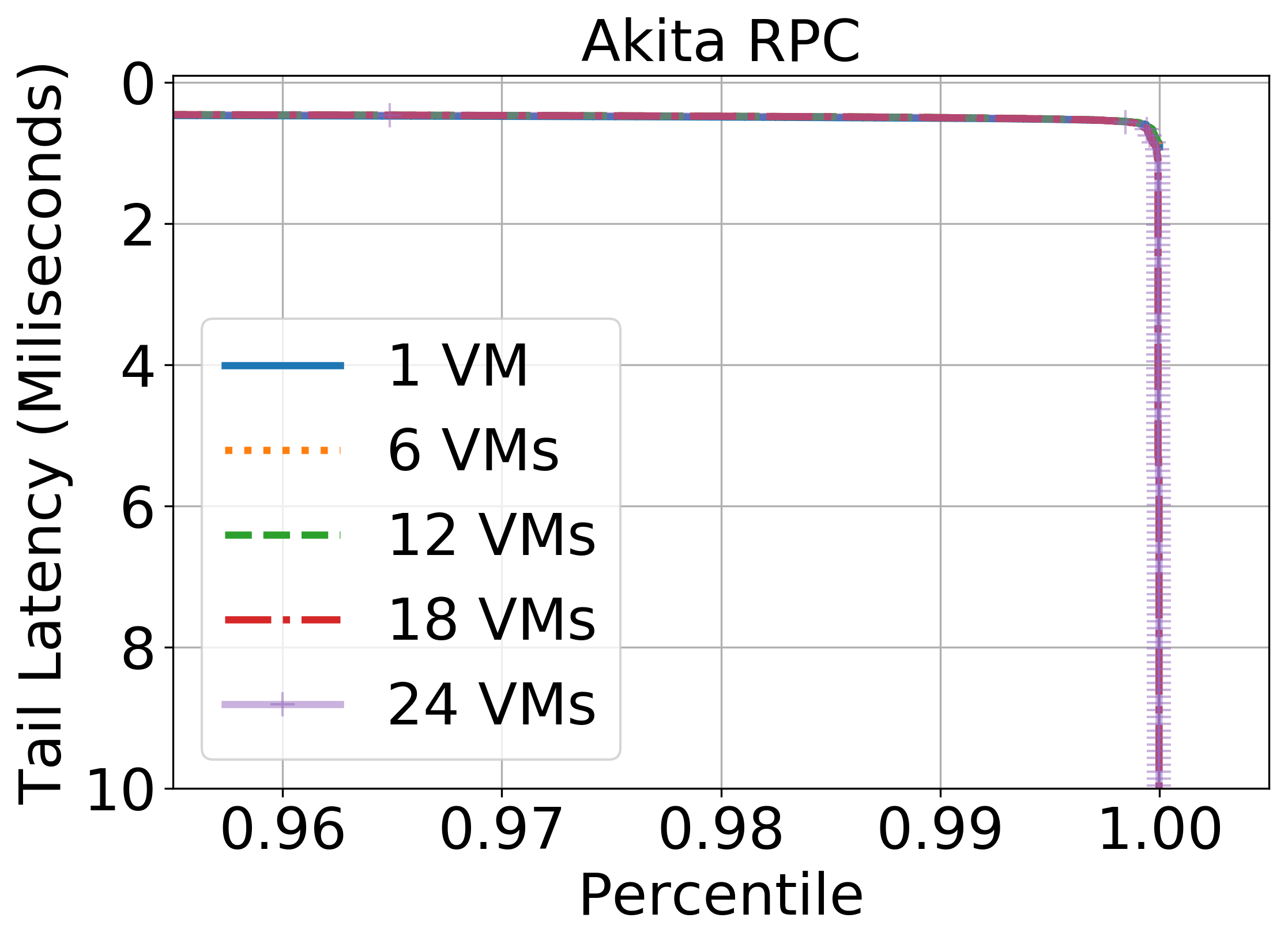}
  \caption{Akita}
  \label{smp1}
 \end{subfigure}
 ~
\begin{subfigure}{0.30\linewidth }
  \centering
  \includegraphics[width=5cm,height=4.5cm, keepaspectratio]{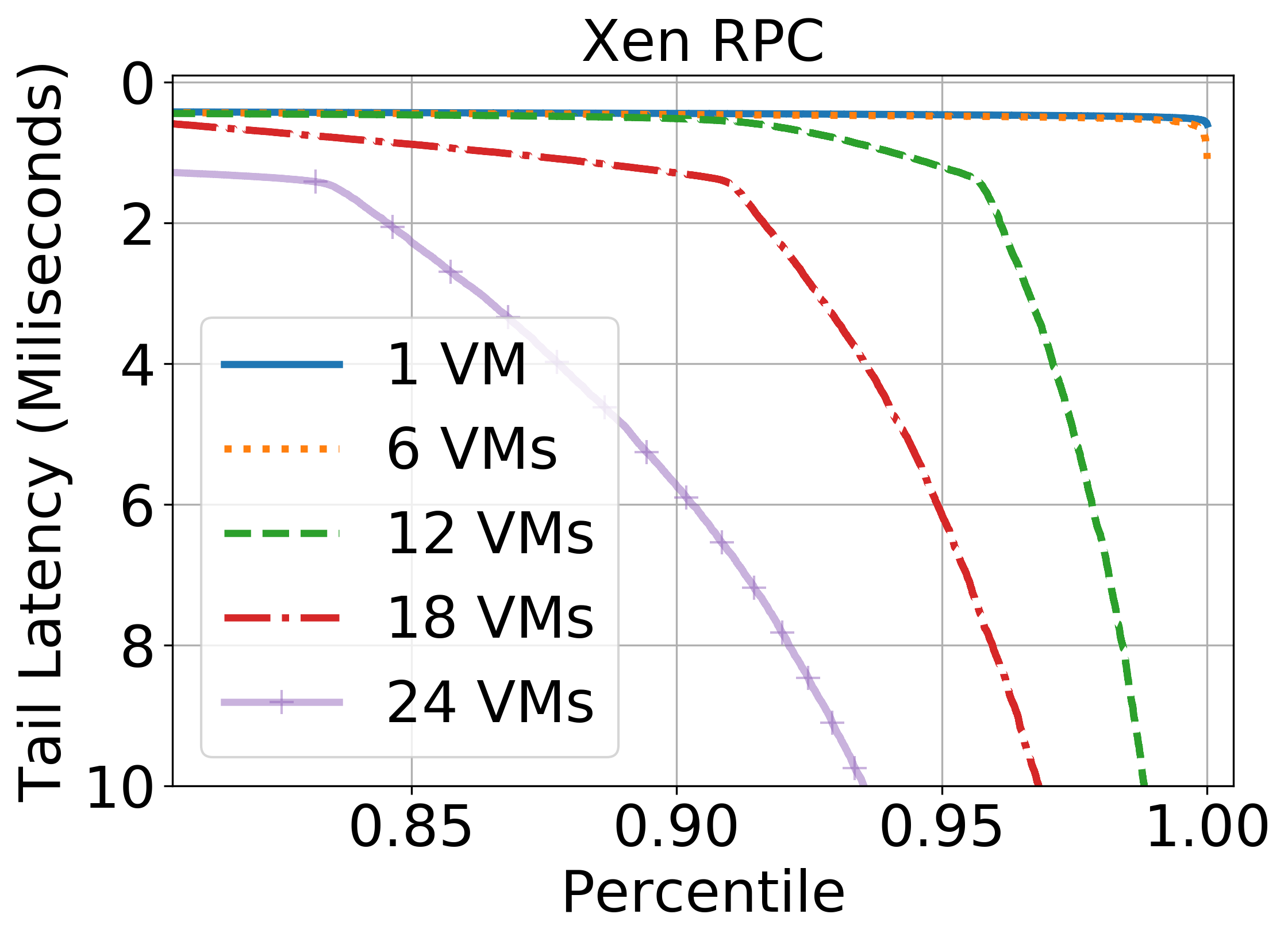}
  \caption{Xen}
  \label{smp2}
\end{subfigure}
~
\begin{subfigure}{0.30\linewidth }
  \centering
  \includegraphics[width= 5cm,height=4.5cm, keepaspectratio]{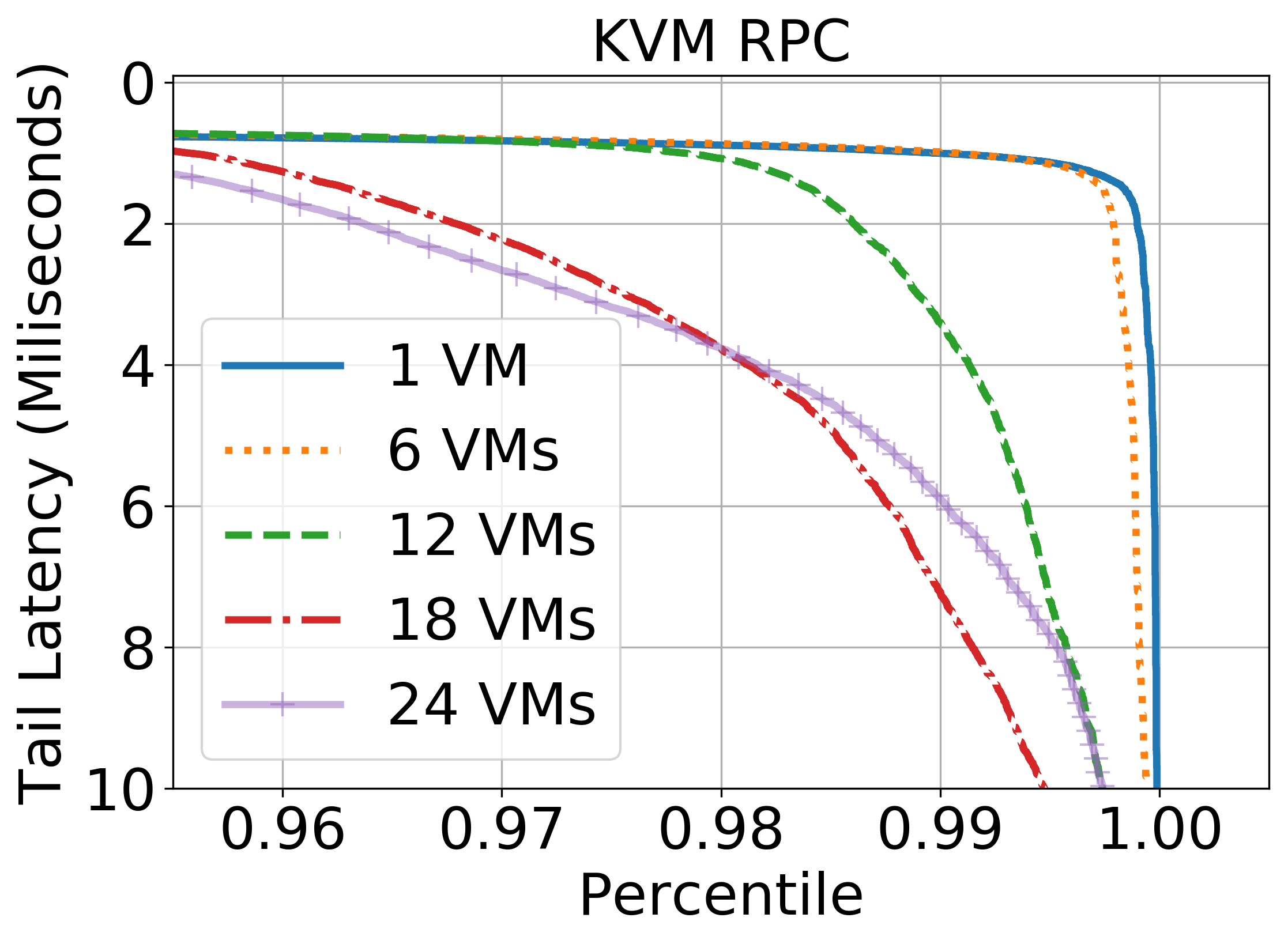}
  \caption{Linux}
  \label{smp3}
\end{subfigure}
\caption{ Tail latency of RPC round trip times }
\label{mem_multi}
\end{figure*}

\begin{figure}
\begin{subfigure}{0.30\linewidth }
  \centering
  \includegraphics[width= 3cm,height=2.5cm, keepaspectratio]{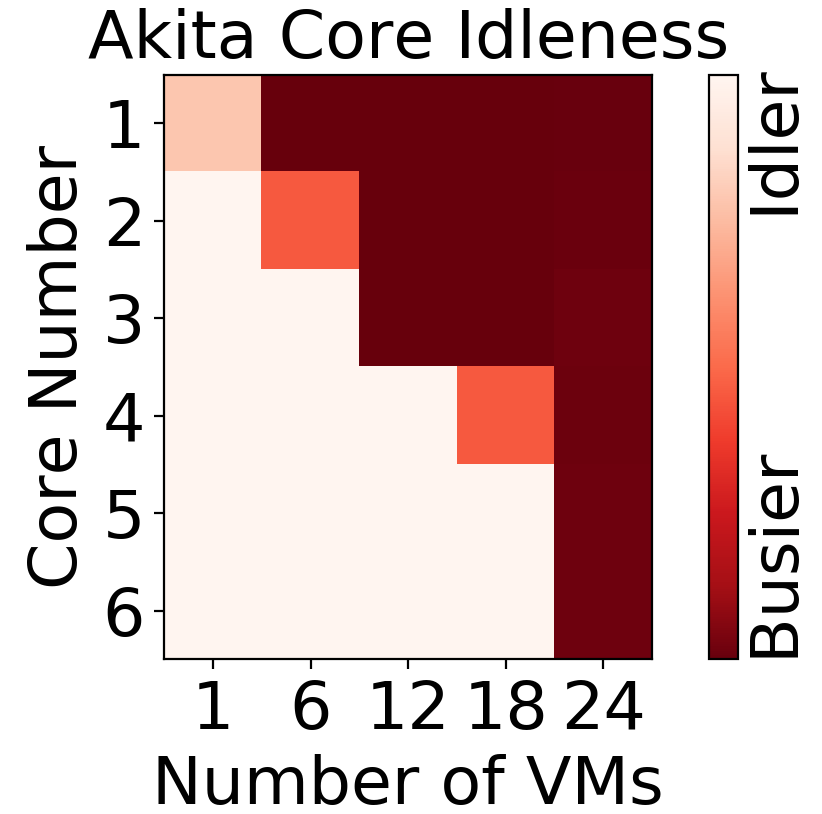}
  \caption{Akita}
  \label{core1}
\end{subfigure}
~
\begin{subfigure}{0.30\linewidth }
  \centering
  \includegraphics[width= 3cm,height=2.5cm, keepaspectratio]{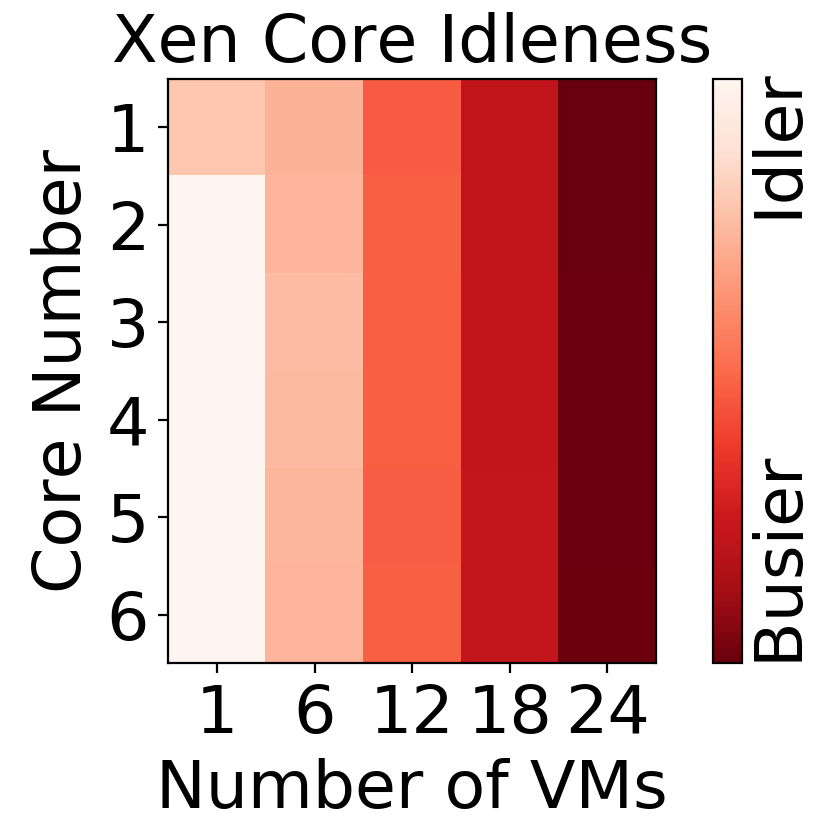}
  \caption{Xen}
  \label{core2}
\end{subfigure}
~
\begin{subfigure}{0.30\linewidth }
  \centering
  \includegraphics[width= 3cm,height=2.5cm, keepaspectratio]{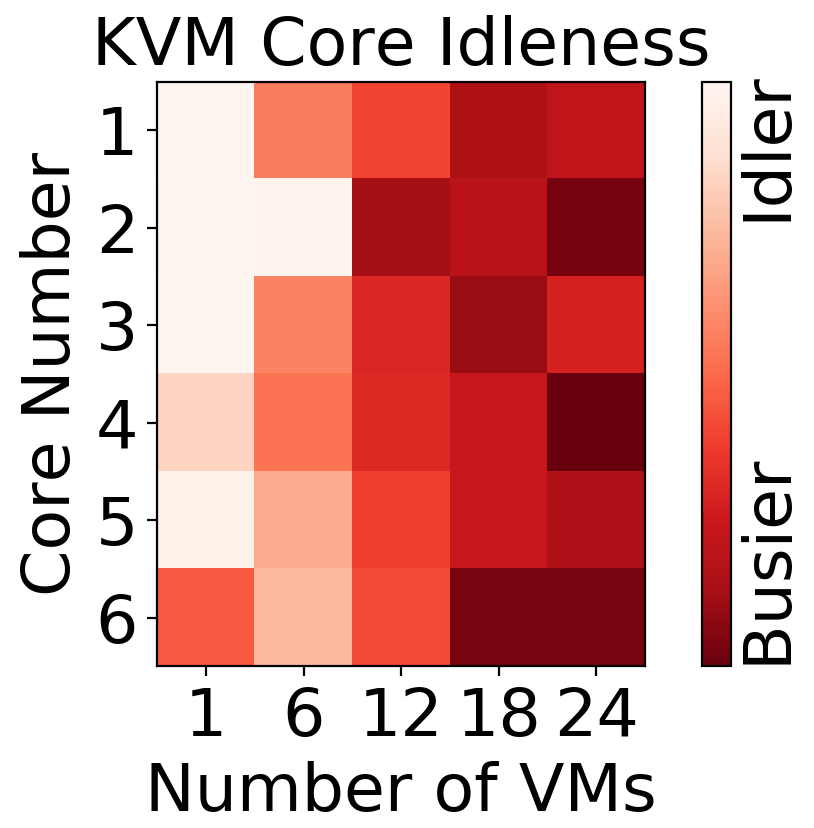}
  \caption{KVM}
  \label{core3}
\end{subfigure}
\caption{ Number of idle-cores under different schedulers }
\label{cores_fig}
\end{figure}

\subsection { Akita's  Mechanism for vCPU Assigning} \label{ssec:smp}
Linux's CFS and Xen's Credit schedulers both leverage a load-balancing mechanism to exploit multicores for raising performance. Under these mechanisms, idle cores steal executable entities from ready queues of busier cores for immediate execution, raising responsiveness of IO-bound vCPUs by mitigating their CALs, and reducing execution times of CPU-bound vCPUs by utilizing more cores.  In contrast, Akita assigns vCPUs to CPU cores using a first-fit mechanism based on the  Akita's schedulability test, does not migrate HI-crit vCPUs, and occasionally migrates LO-crit ones  forced by RQS mechanism.  To compare  best-effort load-balancing policies of Xen and KVM to Akita's approach, we host an RPC server in a VM (Type1) running alongside 5 other CPU-bound VMs (5x Type2 VMs) that consume 25\%  of CPU time and record the average response times of requests sent  from  the client physical machine, each generating 700 calls/second. Note that the RPC VM is a HI-crit VM under Akita. We then increase the number of CPU-bound VMs from 5 VMs up to 24 VMs to utilize more CPU cores.  In Figure ~\ref{mem_multi}, we see that at low loads,  Akita, Xen, and KVM deliver the same performance  because  load balancers of Xen and KVM steal vCPUs from relatively busier cores to  idle cores, shortening  CALs  of the RPC vCPU and hence mitigating RPC RTTs.  This  performance, however, diminishes  as the number of VMs increases. When all cores are utilized (when there are no idle cores), Xen and KVM are not able to hide the negative impacts of collocated VMs on the RPC VM. When 24 VMs run on the physical machine, Xen and KVM lengthen the tail latency at $99^{th}$ percentile by 10x and 5x, respectively.   Akita, on the other hand, delivers a predictable performance by keeping the CALs  of the RPC vCPU under the expected value regardless of the number VMs hosted on the physical machine.  

We report the number of idle CPU cores in Akita, Xen, and Linux during this experiment in Figure ~\ref{cores_fig}. As shown, Akita's first-fit mechanism for assigning vCPUs to cores keeps the number of idle cores and their idleness periods as high as possible.  When  a CPU core does not find any runnable task/vCPU to execute, the operating systems triggers  a mechanism known as idle-state mechanism to turn off processor components in order to reduce  power consumption \cite{Prekas2015-tl} \cite{Kanev2014-yz}.  Increased  number of idle cores in Akita results in more frequent triggering of CPU idle-states, causing more power savings while still delivering the expected performance.  In contrast, Akita's counterparts using their load balancers, spread vCPUs as much as  possible to raise performance, no matter if this level of performance is expected or not,  lowering the number of idle cores and periods of their idleness,  increasing  the energy consumption of CPUs and  thus operational costs of cloud data centers \cite{Gough2015-ty} \cite{Sen2017-jm}.

\subsection {CPU-bound VMs}  \label{ssec:cpu}
In this experiment,  we aim to study if Akita keeps the CPU share of HI-crit VMs intact.   The physical  server machine hosts  VM2, VM3, and VM4 (Type2) that run Lookbusy and each utilizing 25\% of a pCPU.  We also host  VM1 
(Type6)  as a HI-crit VM. We vary CPU utilization of  the HI-crit VM to mimic an unpredictable  CPU-bound  workload. Figure~\ref{cpu} demonstrates how Akita multiplexes the CPU time among these VMs. At first, the LO-crit VMs utilizes  75\%  of CPU time, 25\% each.  When   the load offered to the HI-crit  VM gets intense, the  HI-crit VM  gets its desired CPU share  while the  remaining   CPU time is available to be equally allocated to LO-crit VMs,  suggesting that the  desired CPU share of the HI-crit VM is not impacted by LO-crit VMs which may come at the expanse of unpredictable  utilization of LO-crit VMs. As shown in Figure ~\ref{cpu},  when load offered to the HI-crit VM utilizes  50\%  of the CPU time, each LO-crit VM can only utilize  13\% of the CPU time despite the fact that each one of them needs to utilize 25\% of the CPU time.

\section{Related Work}
\textbf{CPU scheduling in clouds}  is a matter of great concern. Many research works have reported  unpredictable and variable performance of VMs in clouds \cite{vTurbo} \cite{vSlicer} \cite{vPipe} \cite{sort}  \cite{predictability} \cite{Predictable}  \cite{mClock} that stems from traditional policies of CPU scheduling  whose main goal is to raise performance or fairness using best-effort approaches. Xen's round-robin policy \cite{Credit}  \cite{Xen}, for example,   results in  variable CPU access latencies for vCPUs that are responsible for handling IO, hampering responsiveness. vSlicer  \cite{vSlicer} has tacked this challenge by executing latency-sensitive vCPUs more frequently, mitigating CPU access latency (CALs) and thus higher responsiveness. vTurbo  \cite{vTurbo} is another effort that dedicates specific cores to IRQ processing and modifies the guest VMs kernels to schedule IO-bound threads on vCPUs that run on dedicated cores. TerrierTail \cite{terrierTail} reduces the tail latency of VMs running network-intensive workloads by recognizing vCPUs that are receiving network packets and schedules them more frequently, reducing their CPU access latencies and therefore raising responsiveness. 

Although these approaches enhance responsiveness, the delivered QoS is still  not predictable because they try to raise performance of all VMs regardless of VMs SLOs.  Consequently, some approaches suggest avoiding the coexistence of latency- and CPU-intensive VMs on the same physical machine (isolation) to mitigate interference, enhancing predictability at the expense of low utilization. \cite{Bobtail} \cite{Reconciling}. RT-Xen \cite{RT-Xen}  is another approach that aims at delivering real-time performance in virtualized  environments by adopting real-time scheduling in guest OSes and scheduling vCPUs as virtual servers for tasks hosted in guest OSs.  Akita, however, is a cloud CPU scheduler with the aim of delivering predictable IO at high utilization. Although Akita does not require any assumption on guest OS schedulers, by adopting real-time guest OS schedulers, Akita can be used to deliver soft real-time performance in virtualized clouds. Further, Akita offers different level of QoS through  enforcing different CALs  and CPU shares for for both CPU and IO bound workloads. Most importantly, Akita  augments VMs/vCPUs with a criticality-level and discriminate critical VMs from non-critical ones using, allowing the coexistence  of  HI-crit and LO-crit VMs on the same machine and thus raising utilization.

\textbf{ Mixed-criticality scheduling} is a relatively new endeavor that differentiates tasks based on their criticalities for efficient utilization of computing resources while guaranteeing all critical tasks meet their deadlines  \cite{MCS} \cite{Baruah}  \cite{practical}. Vestal initiated MC scheduling by proposing a formalism for a multi-criticality task model and conducting fixed-priority response time analysis for priority assignment \cite{Vestal} \cite{Analysis} . AMC \cite{practical}  is an implementation scheme for uniprocessor scheduling of MC systems. Under AMC, when the system's mode is switched to the HI-crit mode all  LO-crit tasks are abandoned forever. Akita is an SMP scheduler for virtualized clouds. Akita CPUs alternate between HI-crit and LO-crit modes. Akita uses a mode switching mechanism to return back to the LO-crit mode when the conditions are appropriate. 

EDF-VD \cite{EDF} is an MC scheduling algorithm for implicit-deadline sporadic task systems that modifies standard EDF algorithm for the MC priority assignment. EDF-VD shrinks the deadlines of HI-crit tasks proportionally so that the HI-crit tasks will be prompted by the EDF scheduler. Akita leverages EDF-VD to determine the order of the execution of vCPUs. In Akita, unlike EDF-VD, pCPUs in HI-crit mode are reset to LO-crit mode if  HI-crit vCPUs no longer need their  pessimistic CPU shares. Further, Akita reduces power consumption of cloud data centers by using a first mechanism for allocating CPU core while delivering the expected QoS.

\section{Conclusion}
Akita is a  CPU scheduler for virtualized clouds. Akita's main goal is to offer predictable IO even at high utilization of processor resources.   To this end, it first characterizes VMs based on  their CPU  and IO  requirements.  It then categorizes running VMs into HI-crit and LO-crit VMs. Akita ensures a predictable performance  for HI-crit VMs  even when HI-crit VMs are consolidated  with other VMs, which may come at the cost of slowing down the LO-crit VMs temporarily. This allows the coexistence  of HI-crit and LO-crit VMs on the same machine, which notably enhances the utilization of cloud data centers. Experiments with a prototype implementation of Akita demonstrate that a Memcached server VM, as a HI-crit VM, delivers an intact and predictable performance while running alongside several LO-crit CPU-bound VMs.



\bibliographystyle{abbrv}
\bibliography{kv}

\begin{thebibliography}{10}

\bibitem{Credit}
Xen credit scheduler.
\newblock {\em Available: http://wiki.xenproject.org/wiki/Credit\_Scheduler}.
\newblock Last accessed: May 2018.

\bibitem{mClock}
G.~Ajay, A.~Merchant, and P.~J. Varman.
\newblock {mClock: handling throughput variability for hypervisor IO
  scheduling}.
\newblock {\em Proceedings of the 9th USENIX conference on Operating systems
  design and implementation}, 2010.

\bibitem{terrierTail}
E.~{Asyabi}, S.~{SanaeeKohroudi}, M.~{Sharifi}, and A.~{Bestavros}.
\newblock Terriertail: Mitigating tail latency of cloud virtual machines.
\newblock {\em IEEE Transactions on Parallel and Distributed Systems},
  29(10):2346--2359, Oct 2018.

\bibitem{cts}
E.~Asyabi, E.~Sharafzadeh, S.~SanaeeKohroudi, and M.~Sharifi.
\newblock Cts: An operating system cpu scheduler to mitigate tail latency for
  latency-sensitive multi-threaded applications.
\newblock {\em Journal of Parallel and Distributed Computing}, 133:232 -- 243,
  2019.

\bibitem{Xen}
P.~Barham, B.~Dragovic, and K.~Fraser.
\newblock {Xen and the art of virtualization}.
\newblock {\em Proceedings of the Nineteenth ACM Symposium on Operating Systems
  Principles (SOSP)}, pages 164--177, 2003.

\bibitem{MCS}
S.~Baruah, V.~Bonifaci, G.~D'Angelo, H.~Li, A.~Marchetti-Spaccamela, N.~Megow,
  and L.~Stougie.
\newblock Scheduling real-time mixed-criticality jobs.
\newblock {\em IEEE Transactions on Computers}, 61(8):1140--1152, Aug 2012.

\bibitem{Baruah}
S.~Baruah, V.~Bonifaci, G.~DAngelo, H.~Li, A.~Marchetti-Spaccamela, S.~van~der
  Ster, and L.~Stougie.
\newblock The preemptive uniprocessor scheduling of mixed-criticality
  implicit-deadline sporadic task systems.
\newblock In {\em 2012 24th Euromicro Conference on Real-Time Systems}, pages
  145--154, July 2012.

\bibitem{EDF}
S.~K. Baruah, V.~Bonifaci, G.~D'Angelo, A.~Marchetti-Spaccamela, S.~Van
  Der~Ster, and L.~Stougie.
\newblock Mixed-criticality scheduling of sporadic task systems.
\newblock In {\em ESA}, pages 555--566. Springer, 2011.

\bibitem{Analysis}
S.~K. Baruah, A.~Burns, and R.~I. Davis.
\newblock Response-time analysis for mixed criticality systems.
\newblock In {\em 2011 IEEE 32nd Real-Time Systems Symposium}, pages 34--43,
  Nov 2011.

\bibitem{consolidation}
A.~Beloglazov and R.~Buyya.
\newblock {Managing overloaded PMs for dynamic consolidation of virtual
  machines in cloud data centers under quality of service constraints}.
\newblock {\em IEEE Transactions on Parallel and Distributed Systems},
  24(7):1366--1379, 2013.

\bibitem{practical}
A.~Burns and S.~Baruah.
\newblock Towards a more practical model for mixed criticality systems.
\newblock In {\em Workshop on Mixed-Criticality Systems (colocated with RTSS)},
  2013.

\bibitem{Predictable}
R.~C. Chiang, J.~Hwang, H.~H. Huang, and et~al.
\newblock {Matrix: Achieving Predictable Virtual Machine Performance in the
  Clouds}.
\newblock {\em 11th International Conference on Autonomic Computing (ICAC 14)},
  pages 45--56, 2014.

\bibitem{vPipe}
S.~Gamage, C.~Xu, R.~R. Kompellaa, and D.~Xu.
\newblock {vPipe: piped I/O offloading for efficient data movement in
  vrtualized clouds}.
\newblock {\em Proceedings of the ACM Symposium on Cloud Computing}, pages
  1--13, 2014.

\bibitem{Gough2015-ty}
C.~Gough, I.~Steiner, and W.~Saunders.
\newblock {\em {Energy Efficient Servers: Blueprints for Data Center
  Optimization}}.
\newblock Apress, Apr. 2015.

\bibitem{facbook-util}
K.~Hazelwood, S.~Bird, D.~Brooks, S.~Chintala, U.~Diril, D.~Dzhulgakov,
  M.~Fawzy, B.~Jia, Y.~Jia, A.~Kalro, et~al.
\newblock Applied machine learning at facebook: A datacenter infrastructure
  perspective.
\newblock In {\em High Performance Computer Architecture (HPCA), 2018 IEEE
  International Symposium on}, pages 620--629. IEEE, 2018.

\bibitem{PerfIso}
C.~Iorgulescu, R.~Azimi, Y.~Kwon, S.~Elnikety, M.~Syamala, V.~Narasayya,
  H.~Herodotou, P.~Tomita, A.~Chen, J.~Zhang, and J.~Wang.
\newblock Perfiso: Performance isolation for commercial latency-sensitive
  services.
\newblock In {\em 2018 {USENIX} Annual Technical Conference ({USENIX} {ATC}
  18)}, pages 519--532, Boston, MA, 2018. {USENIX} Association.

\bibitem{Jang2015-gf}
K.~Jang, J.~Sherry, H.~Ballani, and T.~Moncaster.
\newblock {Silo: Predictable Message Latency in the Cloud}.
\newblock {\em SIGCOMM Comput. Commun. Rev.}, 45(4):435--448, Aug. 2015.

\bibitem{Kanev2014-yz}
S.~Kanev, K.~Hazelwood, G.~Wei, and D.~Brooks.
\newblock {Tradeoffs Between Power Management and Tail Latency in
  Warehouse-scale Applications}.
\newblock In {\em {2014 IEEE International Symposium on Workload
  Characterization (IISWC)}}, pages 31--40. ieeexplore.ieee.org, Oct. 2014.

\bibitem{Reconciling}
J.~Leverich and C.~Kozyrakis.
\newblock {Reconciling High Server Utilization and Sub-millisecond
  Quality-of-service}.
\newblock In {\em {Proceedings of the Ninth European Conference on Computer
  Systems}}, EuroSys '14, pages 4:1--4:14, New York, NY, USA, 2014. ACM.

\bibitem{predictability}
C.~Li, I.~Goiri, A.~Bhattacharjee, and et~al.
\newblock {Quantifying and improving i/o predictability in virtualized
  systems}.
\newblock {\em IEEE/ACM 21st International Symposium on Quality of Service
  (IWQoS)}, 2013.

\bibitem{OCBP}
H.~Li and S.~Baruah.
\newblock An algorithm for scheduling certifiable mixed-criticality sporadic
  task systems.
\newblock pages 183--192, Nov 2010.

\bibitem{interference}
R.~Nathuji, A.~Kansa, and A.~Ghaffarkhah.
\newblock {Q-clouds: managing performance interference effects for QoS-aware
  clouds}.
\newblock {\em Proceedings of the 5th European conference on Computer systems},
  pages 237--250, 2010.

\bibitem{Elasticity}
R.~R. Nikolas Roman~Herbst, Samuel~Kounev.
\newblock {Elasticity in cloud computing: what it is, and what it is}.
\newblock {\em USENIX 10th International Conference on Autonomic Computing
  (ICAC 2013)}, pages 23--27, 2013.

\bibitem{sort}
D.~Ongaro, A.~L. Coxa, and S.~Rixner.
\newblock {Scheduling I/O in virtual machine monitor}.
\newblock {\em Proceedings of the fourth ACM SIGPLAN/SIGOPS International
  Conference on Virtual Execution Environments}, pages 1--10, 2008.

\bibitem{Shenango}
A.~Ousterhout, J.~Fried, J.~Behrens, A.~Belay, and H.~Balakrishnan.
\newblock Shenango: Achieving high {CPU} efficiency for latency-sensitive
  datacenter workloads.
\newblock In {\em 16th {USENIX} Symposium on Networked Systems Design and
  Implementation ({NSDI} 19)}, pages 361--378, Boston, MA, 2019. {USENIX}
  Association.

\bibitem{Prekas2015-tl}
G.~Prekas, M.~Primorac, A.~Belay, C.~Kozyrakis, and E.~Bugnion.
\newblock {Energy Proportionality and Workload Consolidation for
  Latency-critical Applications}.
\newblock In {\em {Proceedings of the Sixth ACM Symposium on Cloud Computing}},
  SoCC '15, pages 342--355, New York, NY, USA, 2015. ACM.

\bibitem{Arachne}
H.~Qin, Q.~Li, J.~Speiser, P.~Kraft, and J.~Ousterhout.
\newblock Arachne: Core-aware thread management.
\newblock In {\em 13th {USENIX} Symposium on Operating Systems Design and
  Implementation ({OSDI} 18)}, pages 145--160, Carlsbad, CA, 2018. {USENIX}
  Association.

\bibitem{Sen2017-jm}
R.~Sen and D.~A. Wood.
\newblock {Energy-proportional Computing: A New Definition}.
\newblock {\em IEEE Computer}, 2017.

\bibitem{isolation}
D.~Shue, M.~J. Freedman, and A.~Shaikh.
\newblock {Performance isolation and fairness for multi-tenant cloud storage}.
\newblock {\em OSDI'12 Proceedings of the 10th USENIX conference on Operating
  Systems}, pages 349--362, 2012.

\bibitem{utune}
A.~Sriraman and T.~F. Wenisch.
\newblock $\mu$tune: Auto-tuned threading for oldi microservices.
\newblock In {\em 13th USENIX Symposium on Operating Systems Design and
  Implementation (OSDI 18)}, pages 177--194, 2018.

\bibitem{Vestal}
S.~Vestal.
\newblock Preemptive scheduling of multi-criticality systems with varying
  degrees of execution time assurance.
\newblock In {\em 28th IEEE International Real-Time Systems Symposium (RTSS
  2007)}, pages 239--243, Dec 2007.

\bibitem{RT-Xen}
S.~Xi, J.~Wilson, and et~al.
\newblock {RT-Xen: towards real-time hypervisor scheduling in Xen}.
\newblock {\em Proceedings of the IEEE international conference on embedded
  software (EMSOFT)}, page 39–48, 2011.

\bibitem{vTurbo}
C.~Xu, S.~Gamage, and H.~Lu.
\newblock {vTurbo: accelerating virtual machine I/O processing using designated
  Turbo-Sliced core}.
\newblock {\em USENIX ATC'13 Proceedings of the 2013 USENIX conference on
  Annual Technical Conference}, pages 243--254, 2013.

\bibitem{vSlicer}
C.~Xu, S.~Gamage, P.~N. Rao, and et~al.
\newblock {vSlicer: latency-aware virtual machine scheduling via differentiated
  frequency CPU slicing}.
\newblock {\em Proceedings of the 21st ACM International Symposium on
  High-Performance Parallel and Distributed Computing (HPDC '12)}, pages 3--14,
  2012.

\bibitem{Bobtail}
Y.~Xu, Z.~Musgrave, B.~Noble, and M.~Bailey.
\newblock {Bobtail: Avoiding Long Tails in the Cloud}.
\newblock In {\em {NSDI}}, volume~13, pages 329--342, 2013.

\bibitem{Zhuravlev2012-vj}
S.~Zhuravlev, J.~C. Saez, S.~Blagodurov, A.~Fedorova, and M.~Prieto.
\newblock {Survey of Scheduling Techniques for Addressing Shared Resources in
  Multicore Processors}.
\newblock {\em ACM Comput. Surv.}, 45(1):4:1--4:28, Dec. 2012.

\end{thebibliography}

\end{document}